\documentclass[aps,prl,twocolumn,superscriptaddress,longbibliography]{revtex4-1}

\usepackage{mathrsfs}
\usepackage{slashed}
\usepackage{amsmath}
\usepackage{amsfonts}
\usepackage{graphicx,color}
\usepackage{times}
\usepackage[caption=false]{subfig}
\usepackage[colorlinks,linkcolor=red,citecolor=blue]{hyperref}

\AtBeginDocument{%
    \newwrite\bibnotes
    \def\bibnotesext{Notes.bib}
    \immediate\openout\bibnotes=\jobname\bibnotesext
    \immediate\write\bibnotes{@CONTROL{REVTEX41Control}}
    \immediate\write\bibnotes{@CONTROL{%
    apsrev41Control,author="08",editor="1",pages="1",title="0",year="1"}}
     \if@filesw
     \immediate\write\@auxout{\string\citation{apsrev41Control}}%
    \fi
}%

\begin{document}

\title{Exceptional Spin Liquids from Couplings to the Environment}
\author{Kang Yang}
\affiliation{Department of Physics, Stockholm University, AlbaNova University Center, 106 91 Stockholm, Sweden}
\author{Siddhardh C. Morampudi}
\affiliation{Center for Theoretical Physics, MIT, Cambridge MA 02139, USA}
\author{Emil J. Bergholtz}
\affiliation{Department of Physics, Stockholm University, AlbaNova University Center, 106 91 Stockholm, Sweden}

\date{\today}

\begin{abstract}
    We establish the appearance of a qualitatively new type of spin liquid with emergent exceptional points when coupling to the environment. We consider an open system of the Kitaev honeycomb model generically coupled to an external environment. In extended parameter regimes, the Dirac points of the emergent Majorana fermions from the original model are split into exceptional points with Fermi arcs connecting them. In glaring contrast to the original gapless phase of the honeycomb model which requires time-reversal symmetry, this new phase is stable against all perturbations. The system also displays a large sensitivity to boundary conditions resulting from the non-Hermitian skin effect with telltale experimental consequences. Our results point to the emergence of new classes of spin liquids in open systems which might be generically realized due to unavoidable couplings with the environment.
\end{abstract}

\maketitle

Quantum spin liquids are low-temperature phases of matter with fractionalized excitations and emergent gauge fields \cite{anderson1987resonating, balents2010spin, savary2016quantum, Knolle2019, Broholm2020}. Efforts at identifying possible spin liquids has led to hundreds of candidates due to the various possible symmetries present in lattice systems. However, a broader view of the nature of the fractionalized excitations and gauge field leads to only a few prominent types \cite{balents2010spin}, some of which are realized in exactly solvable models \cite{KITAEVToric, KITAEVHoneycomb, PhysRevB.71.045110, yao2007exact, schroeter2007spin, mandal2009exactly, chua2011exact, morampudi2014numerical, buerschaper2014double}. Here, we show that coupling a spin liquid to an environment can lead to a qualitatively new kind of phase which cannot occur in any closed system.

Dissipative systems can display unusual phenomenology not seen in closed systems. These range from unusual phase transitions and critical phases \cite{PhysRevA.87.012108,matsumoto2019continuous,PhysRevLett.123.230401,PhysRevResearch.2.033018} to new topological phases \cite{bergholtz2019exceptional,PhysRevX.8.031079,cerjan2019experimental,Zhou1009}. One prominent class of phenomena can be understood in regimes where a non-Hermitian description \cite{bergholtz2019exceptional,PhysRevX.8.031079,PhysRevLett.121.136802,PhysRevB.99.041406,PhysRevResearch.1.012003,PhysRevLett.123.237202,PhysRevX.9.041015,PhysRevLett.116.133903} of the system is appropriate. This allows for the appearance of exceptional points (EPs) in the spectrum when two eigenvectors coincide \cite{Berry2004,PhysRevX.6.021007,park2020symmetry,Mirieaar7709}. In non-interacting systems, band crossings with such exceptional points result in an unconventional square-root dispersion at low energies as opposed to a typical Dirac dispersion as seen in graphene. These band crossings in 2D systems are generic unlike the accidental symmetry-protected crossings in graphene \cite{bergholtz2019exceptional,Zhou1009}. The conventional bulk-boundary correspondence is also shown to be broken due to an exotic non-Hermitian skin effect \cite{PhysRevLett.121.086803,PhysRevLett.121.026808,Xiong_2018,xiao2020non,PhysRevResearch.2.013058,ghatak2019observation,helbig2020generalized,PhysRevB.99.121101}, which results in localization of all eigenstates at the boundary. This results in an exponential sensitivity of the system to boundary conditions. Based on work of free systems, interest is now drawn to understanding effects in interacting systems \cite{yoshida2019non,lee2020many,matsumoto2019continuous,liu2020non,PhysRevResearch.2.013078,shackleton2020protection,liu2020non,yoshida2020fate}. It is natural to ask what novel features can appear when a system possessing internal strong correlation is described by such an effective non-Hermitian Hamiltonian.

\begin{figure}
    \centering
    \includegraphics[width=0.85\linewidth]{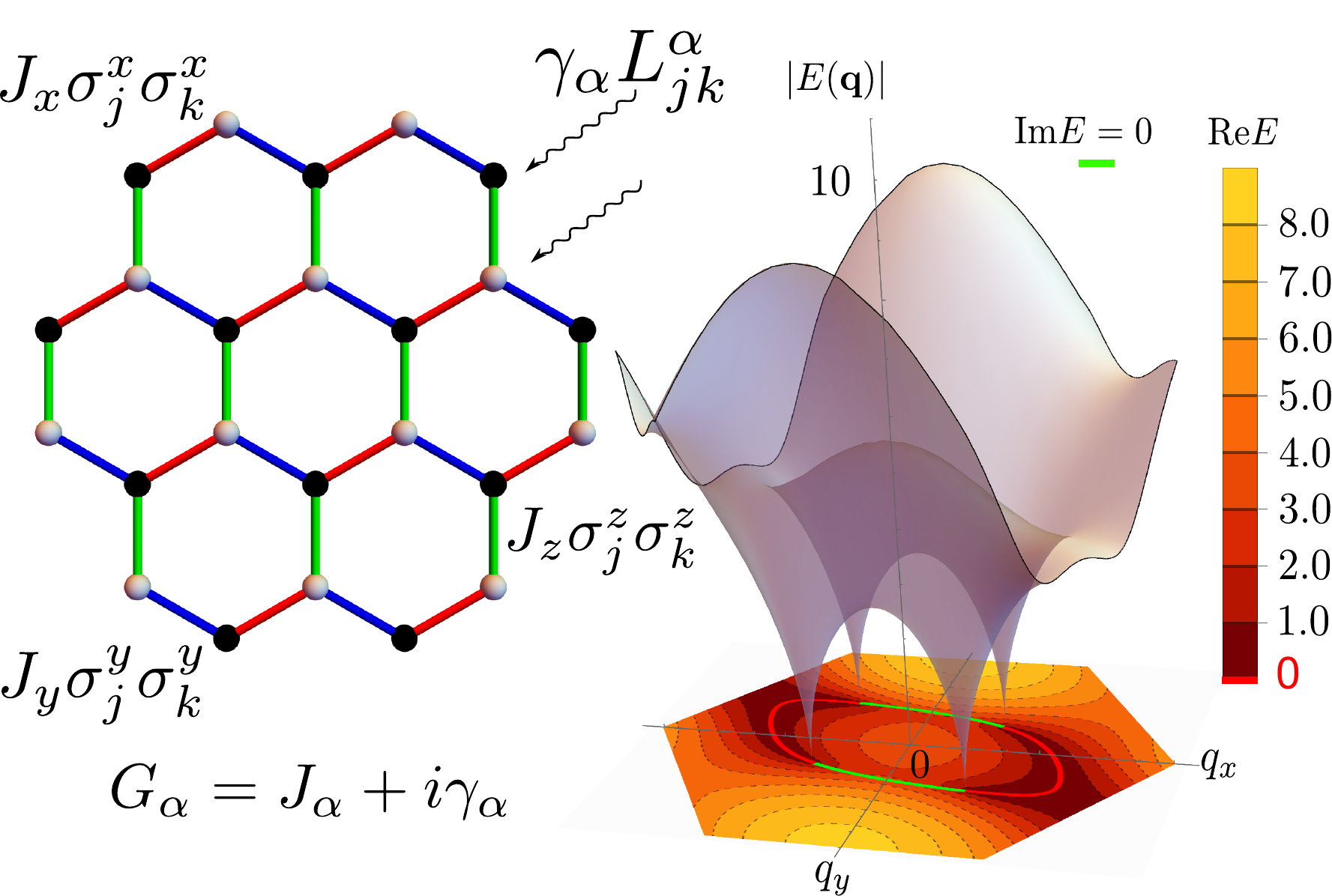}
    \caption{Left: The lattice of the Kitaev honeycomb model. The spins are coupled through $J_\alpha \sigma^\alpha_j\sigma^\alpha_k (\alpha=x,y,z)$. The coupling to the environment is described by jump operators $L^\alpha_{jk}$. Right: the 3D spectrum diagram for the non-Hermitian Kitaev honeycomb model at $G_x=2,G_y=1, G_z=2.5\exp(i\pi/3)$. The Fermi arc ($\textrm{Re}E=0$) is labelled with the red line and the green line indicates the $\textrm{Im}E=0$ curve.}
    \label{fig_3d}
\end{figure}

In this Letter we show that these phenomena can be realized in an interacting spin model giving rise to a qualitatively new kind of spin liquid. We illustrate this by coupling the Kitaev honeycomb model \cite{KITAEVHoneycomb} to an environment (Fig.~\ref{fig_3d}, left panel). In certain regimes, the two Dirac points generically split into four exceptional points (Fig.~\ref{fig_3d}, right panel). The four exceptional points are paired up with each pair connected through Fermi arcs reminiscent to those found in Weyl semimetals \cite{RevModPhys.90.015001}, but occurring in the bulk rather than on the boundaries of the system. Unlike the closed system where the ferromagnetic spin liquid and the anti-ferromagnetic spin liquid are separated by a nodal-line critical point, the open system can go from one to the other by splitting and recombination of exceptional points. Moreover, the only way to produce a gap is to bring the exceptional points together, again similar to Weyl points in 3D. Thus the coupling to the environment elevates a symmetry protected gapless spin liquid to a generically stable phase which we term an {\it exceptional spin liquid}. We also show the occurrence of the skin effect on open zig-zag boundaries leading to a large sensitivity on boundary conditions. Finally, we show that the phenomena are naturally expected to arise in potential realizations of the honeycomb model, such as those proposed in cold atoms and ion traps.

\paragraph{Model}- The Kitaev honeycomb model \cite{KITAEVHoneycomb} is defined through compass interactions linking directions in spin space and real space of spin-$1/2$:
\begin{equation}
    H_0=-\sum_{\langle jk\rangle_\alpha} J_\alpha\sigma^\alpha_j\sigma^\alpha_k,\label{eq_oHam}
\end{equation}
where $\langle jk\rangle_\alpha$ labels the lattice (Fig.~\ref{fig_3d}) and $\alpha=x,y,z$ labelling the three types of links of a hexagonal lattice with $\sigma^\alpha$ the corresponding Pauli matrices. 

We consider an open system where the Kitaev Hamiltonian is coupled to an environment. The resulting open system is described by a Lindblad master equation  \cite{lindblad1976generators} for the density matrix 
\begin{equation}
    \frac{d}{dt}\rho=-i[H_0,\rho]+ \sum_{\langle jk\rangle_\alpha}\gamma_\alpha\left(L_{jk}^\alpha\rho L_{jk}^{\alpha\dagger}-\frac{1}{2}\left\{L_{jk}^{\alpha\dagger} L_{jk}^\alpha,\rho\right\}\right),
    \label{eq:Lindblad}
\end{equation}
where $L_{jk}^\alpha$ are jump operators describing how the system is coupled to the bath and we have set $\hbar=1$. The dynamics can be interpreted in terms of deterministic evolution of a trajectory (wavefunction) described by an effective non-Hermitian Hamiltonian $H_{\rm NH}=H_0-(i/2)\sum_{\langle jk\rangle_\alpha} \gamma_\alpha L^{\alpha\dagger}_{jk} L_{jk}^\alpha$, interspersed with quantum jumps to different states through the $L_{jk}^\alpha\rho L_{jk}^{\alpha\dagger}$ term \cite{dum1992monte, molmer1993monte, plenio1998quantum, Daley2014}. Thus when we are measuring at times before the first jump, the dynamics is governed by the non-Hermitian Hamiltonian $H_{\rm NH}$. This can be interpreted as a measurement backaction in settings where realizations of the model system are post-selected to consider only cases where the jump has not occurred \cite{plenio1998quantum, Daley2014, lee2014heralded, ashida2016quantum, nakagawa2020dynamical}. 

Although the general phenomenology of what follows is largely independent of the form of the jump operators, we consider here jump operators $L_{jk}^\alpha=\sigma_j^\alpha+\sigma_k^\alpha$ along each each $\alpha$-type link $j-k$ for illustration. Similar results can be obtained by considering the effect of dephasing noise \cite{PhysRevB.99.174303}. This results in an effective  non-Hermitian description of the the form of Eq.~(\ref{eq_oHam}) but with the coupling constants being complex and henceforth labeled by $G_\alpha=J_\alpha+i \gamma_\alpha$. This model is exactly solved through an enlarged Majorana representation of the spin operators. Introducing four Majorana fermions $(c_{j}, b_{j}^{x}, b_{j}^{y}, b_{j}^{z})$ at each site, the spin is represented as $\sigma_{j}^{\alpha}=i c_{j} b_{j}^{a}$. The physical state is constrained by the condition: $D_j|\textrm{phys}\rangle=|\textrm{phys}\rangle$, with $D_j=b^x_jb^y_jb^z_jc_j$. Defining bond operators $u_{jk}=i b_{j}^{\alpha} b_{k}^{\alpha}$, where $\alpha$ is the type of link $j-k$, the effective model is 
\begin{equation}
H_{\rm NH}=-\sum_{\langle jk\rangle_\alpha} G_\alpha\sigma^\alpha_j\sigma^\alpha_k=i \sum_{\langle jk\rangle_\alpha} G_{\alpha} u_{jk} c_{j} c_{k}\ .\label{HnH}
\end{equation}
The bond operator $u_{jk}$ is a constant of motion in the enlarged Majorana representation. Meanwhile, a physical conserved quantity is the product of $u_{jk}$ around a plaquette $p$, $ W_p=\prod_{j-k\in p} u_{jk}$. The eigenvectors can thus be grouped into different sectors of the eigenvalues $W_p=\pm 1$. The sector with all $ W_p=1$ is viewed as vortex free and $W_p=-1$ means a $\mathbb Z_2$-vortex at $p$.

\begin{figure*}
    \centering
    \includegraphics[width=0.85\linewidth]{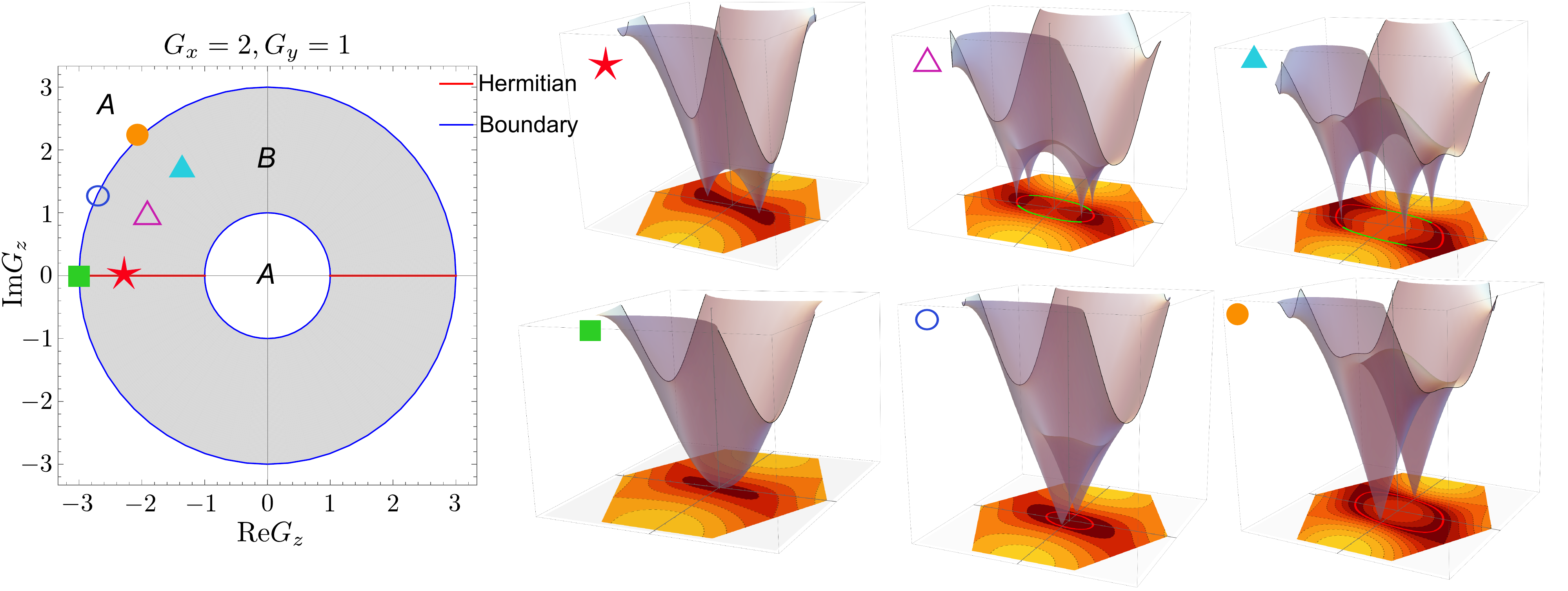}
    \caption{The spectrum and phase diagram for the non-Hermitian Kitaev honeycomb model at $G_x=2,G_y=1$. The grey region $B$ admits band-touching points while the region $A$ possesses a (complex) gap. Six values of $G_z$ are taken to exhibit different dispersions with respect to $\mathbf q$ as in Fig.~\ref{fig_3d}. For the Hermitian situation (the star), around the band-touching point $E\sim \sqrt{g_{\alpha\beta}\delta q_\alpha\delta q_\beta}$ with coefficients $g_{\alpha\beta}$. At the Hermitian phase boundary (the square), the dispersion is quadratic along a certain direction $E\sim \sqrt{(g_\alpha \delta q_\alpha)^4}$ with coefficients $g_{\alpha}$. In the non-Hermitian situation (the two triangles), the dispersion is square-root $E\sim \sqrt{g^\prime_\alpha\delta q_\alpha}$. At the non-Hermitian boundary (the two circles), the energy is linear along some direction $E\sim \sqrt{g^\prime_{\alpha\beta}\delta q_\alpha\delta q_\beta} $.}\label{fig_phdia}
\end{figure*}

The eigenstates of the model can be decomposed into different $\mathbb Z_2$-flux sectors as in the original Hermitian model, where the zero-flux is the relevant one at low (real) energies due to Lieb's theorem \cite{KITAEVHoneycomb,PhysRevLett.73.2158}. In fact, the zero-flux sector is still relevant for the open systems in appropriate regimes where all the phenomenology we discuss is realized. To illustrate this, consider the Hermitian model at temperatures much lower than the vison gap. If we consider multiplying it by an overall complex number, there is a parametric separation in lifetimes of states between different flux sectors and the zero-flux sector corresponds to states with the longest lifetimes. In the supplementary material we provide numerical evidence that this holds true also in more general cases \cite{suppm}. 

In the zero-flux sector we can chose $u_{jk}=1$ for $j$ on one sublattice and $k$ on the other. The Hamiltonian becomes a tight binding Majorana model, in momentum space as:
\begin{equation}
    \widetilde H=\sideset{}{'}\sum_{\mathbf q}\left(\begin{array}{ccc}
         c_{-\mathbf q,1} &
         c_{-\mathbf q,2} 
    \end{array}\right)\left(\begin{array}{ccc}
        0 & iA(\mathbf q)\\
        -iA(-\mathbf q) & 0
    \end{array}\right)\left(\begin{array}{ccc}
         c_{\mathbf q,1} \\
         c_{\mathbf q,2} 
    \end{array}\right),\label{eq_mHmtn}
\end{equation}
where in the primed summation $\sideset{}{'}\sum$ we count the pair $\mathbf q,-\mathbf q$ only once. This is because $c_{-\mathbf q}=c^\dagger_{\mathbf q}$. The off-diagonal element is $A(\mathbf q)=2\left(G_xe^{i\mathbf q\cdot\mathbf r_1}+G_ye^{i\mathbf q\cdot\mathbf r_2}+G_z\right)$ and the subscripts $1,2$ label the two sublattices of the honeycomb system. The momentum-space operators satisfy $\{c_{\mathbf q,\lambda},c_{-\mathbf q',\gamma}\}=\delta_{\mathbf q,\mathbf q'}\delta_{\lambda, \gamma}$. Depending on whether $A(\mathbf q)$ can go to zero, the system exhibits a gapped or a gapless phases. In the Hermitian model, the gapped phase is equivalent to a toric code spin liquid \cite{KITAEVToric} while the gapless phase can possess non-Abelian statistics in the presence of magnetic field \cite{KITAEVHoneycomb}.  The gapless condition is given that the lengths $|G_x|,|G_y|,|G_z|$ admit a triangle:
\begin{equation}
    |G_x|\le |G_y|+|G_z|,\ |G_y|\le |G_x|+|G_z|,\ |G_z|\le |G_x|+|G_y| \label{eq_gapless}.
\end{equation}

\paragraph{Exceptional points and Fermi arcs}- The spectrum is obtained by the eigenvectors of the matrix  \eqref{eq_mHmtn} $E^2(\mathbf q)=A(\mathbf q)A(-\mathbf q)$. In the Hermitian case, there are two Dirac points inside the gapless region. We show that for complex $G_\alpha$, the band touching points have a square-root dispersion and are exceptional points.

For convenience of computation here, we can extract out the phase of $G_z$ as an overall phase of $\widetilde H$, so the Hamiltonian is now parameterized by $\bar \phi_x=\phi_x-\phi_z,\bar\phi_y=\phi_y-\phi_z$. In the Brillouin zone, it is more convenient to parametrize as $\mathbf q=\mathbf q_1\tilde q_1/(2\pi)+\mathbf q_2\tilde q_2/(2\pi)$, where $\mathbf q_1$ and $\mathbf q_2$ are the reciprocal lattice vectors. The values $\tilde q_1,\tilde q_2$ uniquely fix the vector $\mathbf q$ (notice that we should take $\tilde q_1,\tilde q_2\textrm{ mod } 2\pi$). Zero energy at $\mathbf q$ implies $A(\mathbf q)=0$ or $A(-\mathbf q)=0$. The $A(\mathbf q)=0$ condition gives $\tilde q_1,\tilde q_2$
\begin{equation}
\tilde q_{1(2)}=\pm\cos^{-1}\left(\frac{|G_{y(x)}|^2-|G_z|^2-|G_{x(y)}|^2}{2|G_{x(y)}||G_z|}\right)-\bar\phi_{x(y)} 
\end{equation} 
with the constraint $G_x\sin(\tilde q_1+\bar\phi_x)=-G_y\sin(\tilde q_2+\bar\phi_y)$ to fix the $\pm$ signs above. These equations admit at most two solutions which we denote them as $\mathbf q_e, \mathbf q'_e$. In the Hermitian situation, one has $A^*(\mathbf q)=A(-\mathbf q)$ and $\mathbf q_e=-\mathbf q'_e$ as $\bar\phi_x=\bar\phi_y=0$. Linearizing $A$ it directly follows that we have Dirac points with a conventional linear dispersion away from the degeneracies.

For complex, i.e., non-Hermitian, parameters, we have $A^*(\mathbf q)\neq A(-\mathbf q)$ and $\mathbf q_e=-\mathbf q'_e-2(\bar\phi_x,\bar\phi_y)$. Now $A(\mathbf q)$ and $A(-\mathbf q)$ vanish at different points, implying four $E=0$ exceptional points $\pm\mathbf q_e,\pm\mathbf q'_e$, at which the Hamiltonian matrix becomes non-diagonalizable and the two eigenvectors coincide. The dispersion near the exceptional points takes a square-root form instead of the conic form since $A(\mathbf q)$ and $A(-\mathbf q)$ do not vanish simultaneously. The $\textrm{Re} E=0$ branch cuts associated with the square-roots have a natural interpretation as bulk Fermi arcs, and the exceptional points are connected by these Fermi arcs and their imaginary counterparts $\textrm{Im} E=0$ (cf. Fig.~\ref{fig_3d}, right panel). 

When the coupling constants are tuned out of the triangle regime Eq.~\eqref{eq_gapless}, the four exceptional points fuse into two exceptional points and then disappears. A cut at $G_x=2,G_y=1$ for different complex $G_z$ is shown in Fig.~\ref{fig_phdia}, where we also plot the absolute energy $|E|$ and its real part, ${\rm Re}E$, at different $G_z$. One can see the splitting of each Hermitian band-touching point into two non-Hermitian exceptional points. At the phase boundary, the branch-cut for $E$ disappears. In this situation, the band-touching point is not protected and thus gets gapped out when crossing the phase boundary. This illustrates how, similar to Weyl points in 3D, the exceptional points can only be gapped out when combined pairwise, in glaring contrast to the 2D Dirac points that are inherently symmetry protected.

\begin{figure*}
     \centering
     \subfloat[]{
     \includegraphics[width=0.29\linewidth]{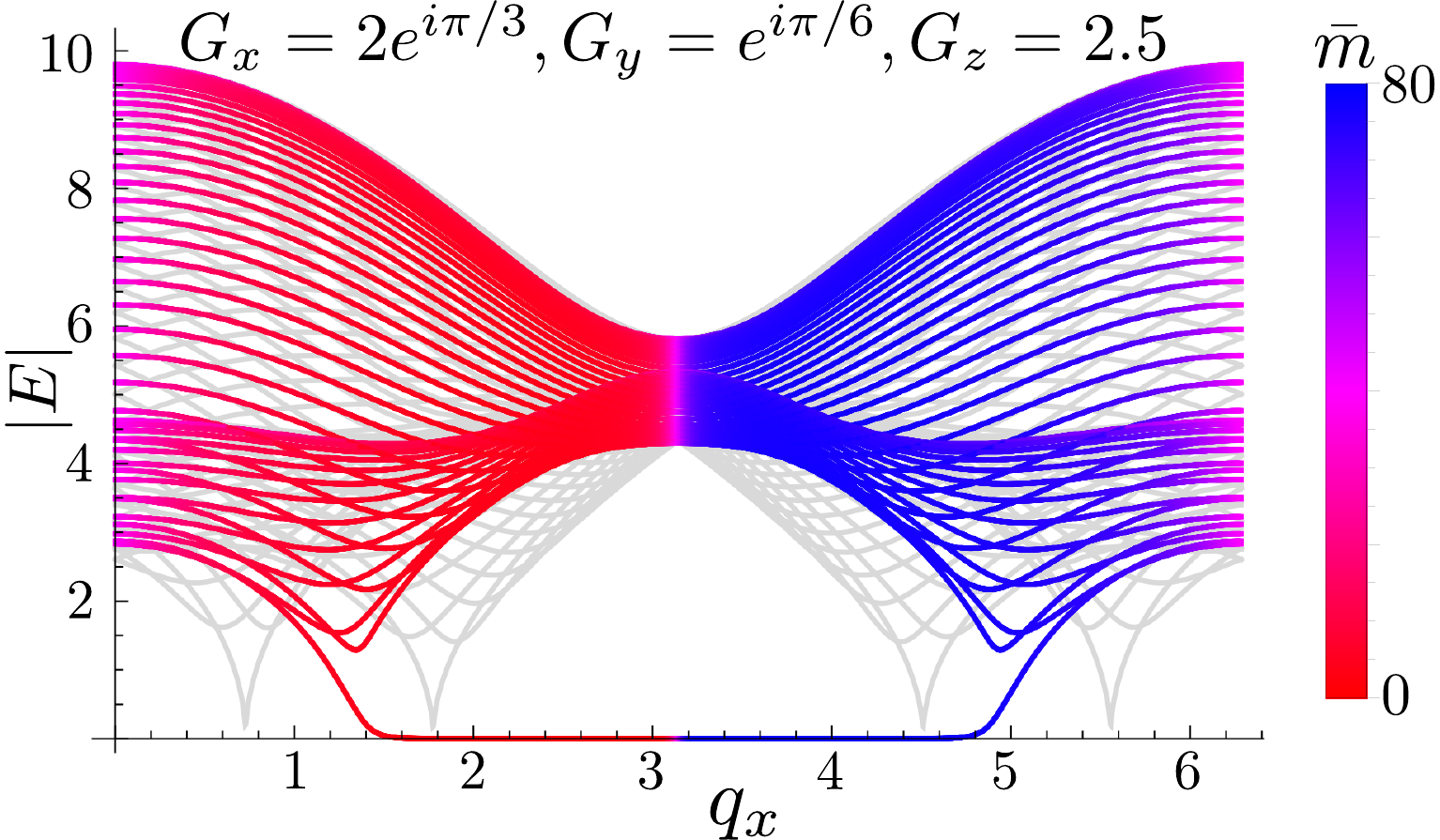}
    \label{fig_opbspec}}
     \subfloat[]{
    \includegraphics[width=0.29\linewidth]{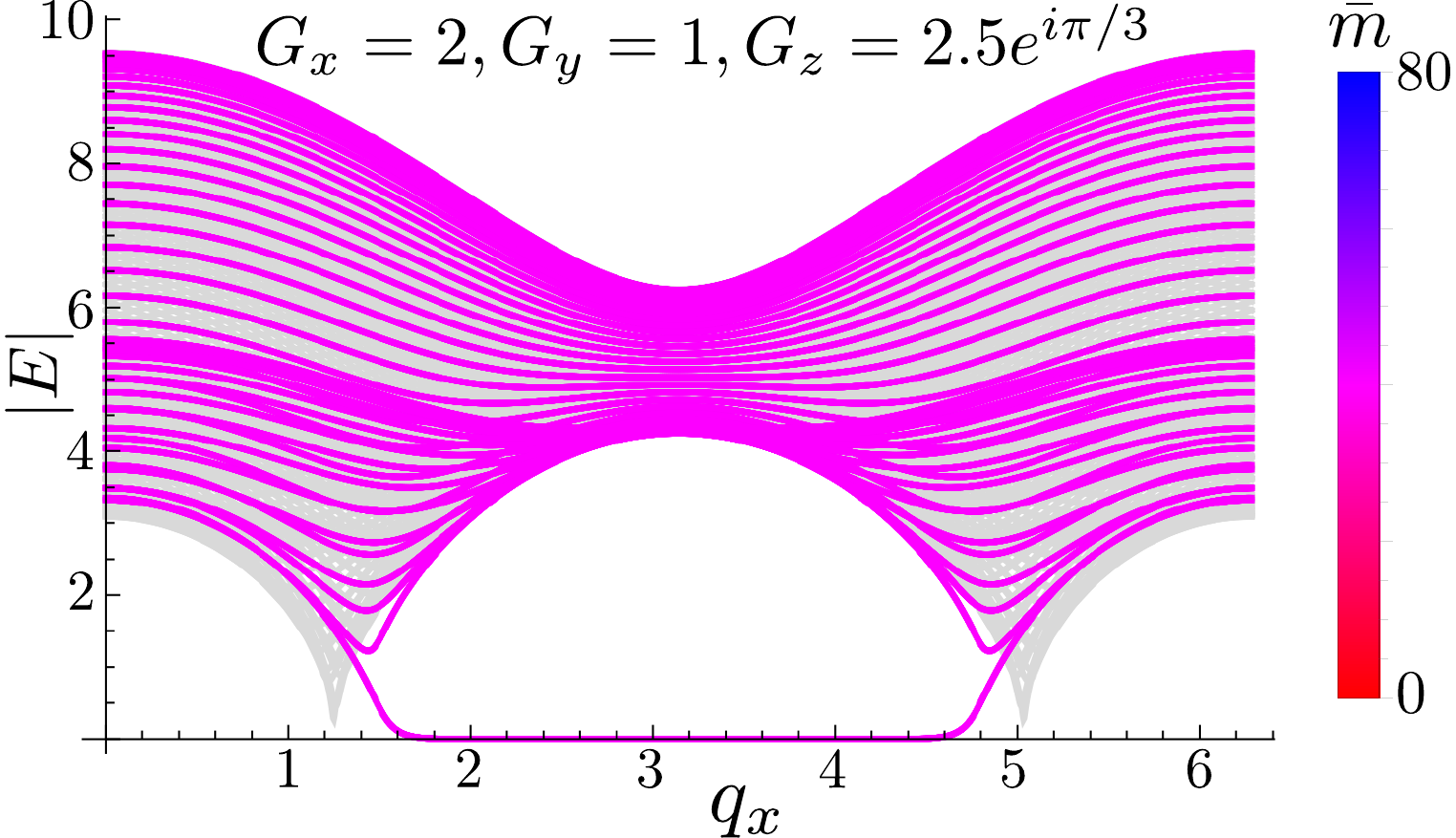}
    \label{fig_skineffN}
   }
     \subfloat[]{
    \includegraphics[width=0.29\linewidth]{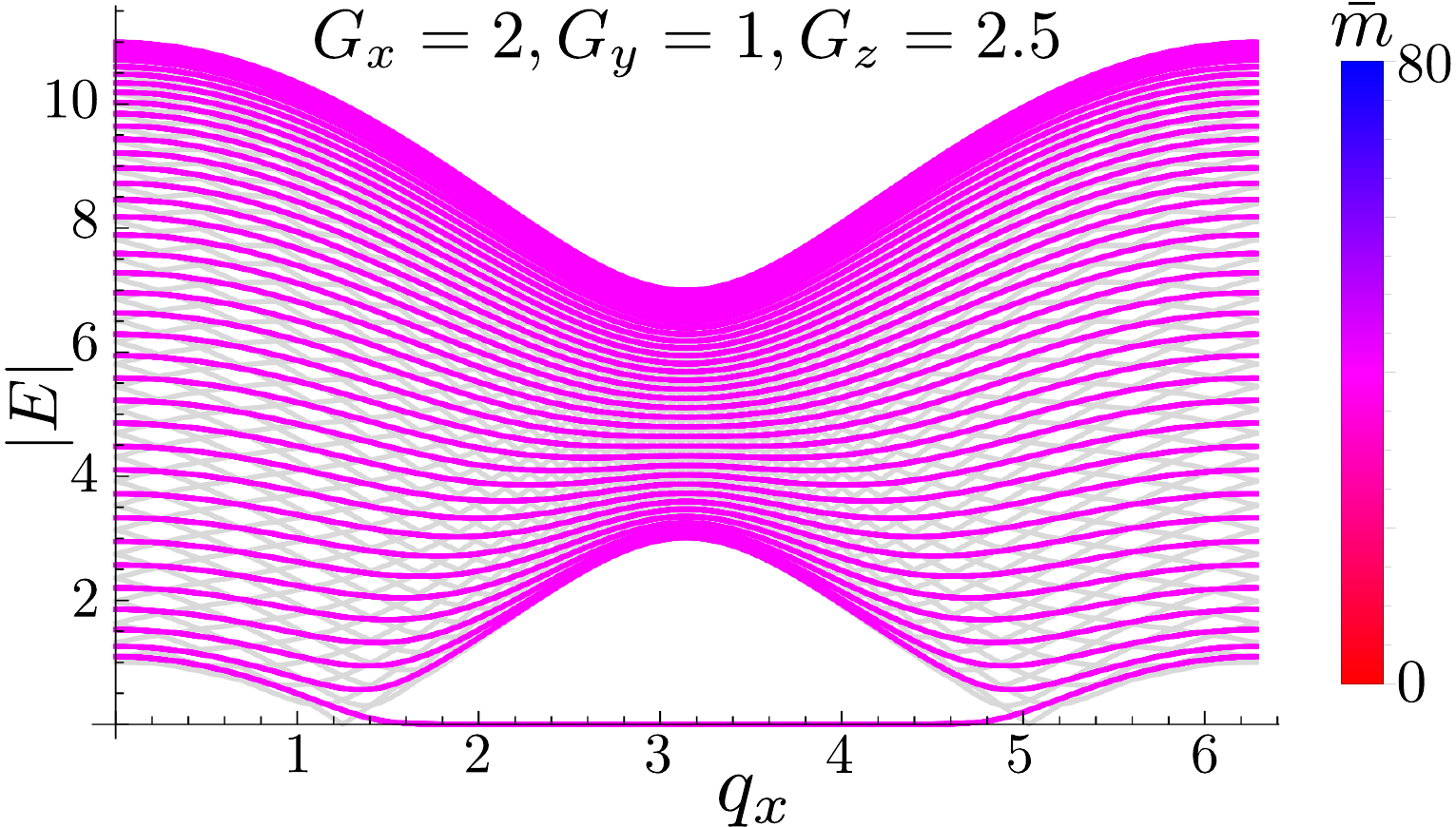}
    \label{fig_opbspech}
    }
    
    \caption{The spectra of the system for OBC (colored) and  PBC (light gray) and the corresponding average localization of the eigenstates. (a) A non-Hermitian skin effect occuring due to the non-trivial relative phase $\phi_x-\phi_y\ne 0\  (\textrm{mod }\pi)$. (b) The skin effect does not occur if only $G_z$ is tuned complex while $\phi_x-\phi_y= 0\  (\textrm{mod }\pi)$. (c) The Hermitian case, where the PBC and OBC spectra overlap except for the edge mode. All results are obtained for a $M=80$-row lattice.}
\end{figure*}

\paragraph{Skin effects}- Boundary conditions do not affect the spectrum for Hermitian systems in the thermodynamic limit except for additional edge states. In contrast, non-Hermitian systems display a strong sensitivity to boundary conditions. To illustrate this, we consider Eq.~(\ref{HnH}) with two parallel zigzag boundaries. We place the open boundary condition (OBC) perpendicular to the $z$-type link, which is along the $y$-direction. The $x$-direction is still chosen to be periodic with $N_x$ unit cells. For convenience of comparison with the periodic boundary condition (PBC), the number of layers sandwiched by the two boundaries is taken to be even $M=4M'$. The Hamiltonian is diagonal in the momentum $q_x$ and we denote the state by $\psi(m)$ with $m$ the layer index. The boundary conditions require  $\psi(0)=0$ and $\psi(M+1)=0$. The question can be solved by a transfer matrix method \cite{PhysRevB.99.245116,suppm}. We group the wave functions into a doublet $\Psi(m)=(\psi(2m),\psi(2m+1))$. Then the equation of motion takes the form $\Psi(m)=T\Psi(m-1)$, where $T$ is the transfer matrix (details in \cite{suppm}). Each eigenstate $\Psi(m)$ can be therefore constructed as a superposition of the two eigenvectors of $T$, $\Psi(m)=\alpha s_1^m\Psi_1+\beta s_2^m\Psi_2$ with $\Psi_{1,2}$ the eigenvectors of $T$ and $s_{1,2}$ the corresponding eigenvalues, such that it satisfies $\psi(0)=\psi(M+1)=0$. If we consider Hermitian couplings, then $|s_1|=|s_2|= 1$ and the eigenstates propagate into the bulk. However in the non-Hermitian situation, $s_1\simeq s_2$ are either both larger or smaller than one and the states are piled up against one of the boundaries. This is known as the non-Hermitian skin effects \cite{PhysRevLett.121.086803,PhysRevLett.121.026808}. The general criterion for the skin effect is when $|s_1s_2|$ given by 
\begin{equation}
    |\det T| =\frac{\left|G_xe^{iq_xa}+G_y\right|}{\left|G_xe^{-iq_xa}+G_y\right|}\neq 1.\label{eq_ratioT}
\end{equation}
In this case, all eigenstates are exponentially localized to the boundary $\psi(m)\sim \exp(-m/l)$, with $l=(\ln |\det T|)/2$. We see that this occurs when the relative phase $\phi_x-\phi_y$ is non-zero. By rotation symmetry, we can draw the conclusion that for two parallel zigzag open boundaries perpendicular to $\alpha$-type links, the skin effects can be turned on by giving a non-trivial relative phase $\phi_\beta-\phi_\gamma$, where $\beta,\gamma\ne \alpha$. 

In Fig.~\ref{fig_opbspec}-\ref{fig_opbspech}, we show the OBC spectra for different constants. The average localization of the wave function, $\bar m=\sum m|\psi(m)|^2$, is indicated with the color plot. For a non-vanishing $\phi_x-\phi_y$, as in Fig.~\ref{fig_opbspec}, in addition to the zero-energy boundary state, the bulk states are also piling up the $m=1$ boundary, exhibiting the skin effects. The localization shift from one boundary to the other at $q_x=0,\pi$, in accordance with Eq.~\eqref{eq_ratioT}. The spectrum is strikingly different from the PBC spectrum. For a vanishing $\phi_x-\phi_y$ and non-vanishing $\phi_z$, we can however see in Fig.~\ref{fig_skineffN} that there is no skin effect despite the presence of bulk exceptional points and the OBC spectrum coincide with the PBC spectrum. Fig.~\ref{fig_opbspech} shows a Hermitan example contrasting the novel non-Hermitian behaviour.

\paragraph{Discussion}- We have shown in this Letter that genuinely non-Hermitian phenomenology, exceptional points and skin effects, intriguingly conspire with fractionalization in the interacting Kitaev honeycomb model in dissipative environments. This results in a qualitatively new type of non-equilibrium matter which we call {\it exceptional spin liquids}, which is by its dissipative nature, lies beyond earlier classification schemes and potentially displays new dynamics beyond current spin liquids \cite{knolle2014dynamics, morampudi2017statistics, werman2018signatures, morampudi2020spectroscopy}. Remarkably, this new phase is generic in the sense that it does not rely on any underlying symmetries---this may in fact greatly facilitate the prospects for observing gapless spin liquids in synthetic setups. 

The exceptional points can naturally arise in many of the proposed realizations of the Kitaev honeycomb model \cite{Duan2003, Micheli2007, Jackeli2009, schmied2011quantum, you2010quantum, gorshkov2013kitaev}. Two concrete mechanisms involve incorporating effects of a self-energy \cite{yoshida2020exceptional} and a post-selection procedure \cite{plenio1998quantum, Daley2014, ashida2016quantum, naghiloo2019quantum} on synthetic realizations. The material candidates for the Kitaev honeycomb model could potentially realize this non-Hermitian phenomenology if the effects of excitations such as phonons are taken into account. Here, we expound on the post-selection procedure in a synthetic optical lattice systems of ultracold atoms. Such systems are unavoidably subject to dissipative effects, most notably from spontaneous emission and inelastic collisions \cite{bloch2008many, riou2012spontaneous}. As mentioned above (Eq.~\ref{eq:Lindblad}), the absence of quantum jumps from such processes leads to a measurement backaction \cite{plenio1998quantum, Daley2014, lee2014heralded, ashida2016quantum, nakagawa2020dynamical} resulting in a non-Hermitian evolution. If spontaneous emission is the dominant dissipative mechanism, the procedure is implemented by a continuous measurement of the system to monitor for emitted photons and a post-selection on those realizations where no photons are observed.

Considering ${}^{87}\mathrm{Rb}$ atoms with a typical magnetic scale of 100 Hz, the scattering rate due to spontaneous emission is a few Hz. It is dependent on the detuning $\Delta$ of the laser as $\Delta^{-2}$ and can hence be tuned appropriately across a wide range from milliseconds to many minutes \cite{friebel1998co}. Further tuning can be achieved by using different transitions in the same or different atoms since the spontaneous emission rate goes as $\omega_0^3$ where $\omega_0$ is the transition energy. The losses due to inelastic collisions can also be tuned through techniques such as external confinement, Feshbach resonance and photoassociation \cite{bloch2008many}. The Kitaev honeycomb model can be realized with hyperfine states and a spin-dependent potential \cite{Duan2003}. Using different detunings for the different lasers creating the lattice can then result in the direction-dependent Lindblad operators necessary for the phenomenology here. Although it would seem beneficial to have a detuning which is as large as possibles since that would lead to a bigger separation of the exceptional points, it is practically limited by the transition frequency. More importantly, it also make the post-selection procedure more difficult since the Poissonian decay process means that the probability of a decay not occuring would be $\sim e^{-N\gamma/\delta}$ where $\gamma$ is the noise rate, $N$ is the number of spins and $\delta$ is the imaginary part of the many-body gap. Hence, the primary challenge would be to resolve the exceptional points due to a small separation. Choosing parameters such that the zero-flux sector has the longest lifetime (see Supplementary) would give an enhancement to the naive lifetime and might allow one to choose larger noise rates to better resolve the exceptional points.  

The exquisite control and ubiquitous presence of dissipation in the suggested synthetic implementations of our ideas might even open the door for novel technological applications such ultra sensitive sensing devices based on harnessing the non-Hermitian skin effect \cite{budich2020non} by judiciously manipulating the boundary conditions. 

As strongly correlated many-body states exhibit a rich variety of emergent phenomena, such as non-Abelian statistics, the study of their interplay with genuinely non-Hermitian effects as advanced here is likely to provide fertile ground for new fundamental discoveries.  
\\
\acknowledgments{\paragraph{Acknowledgements}- 
KY and EJB acknowledge funding from the Swedish Research Council (VR) and the Knut and Alice Wallenberg Foundation. SM acknowledges funding from the Tsung-Dao Lee Institute. 
} 
\bibliography{NHK}

\begin{widetext}

\section{Supplemental Material}
\renewcommand{\theequation}{S\arabic{equation}}
\setcounter{equation}{0}
\renewcommand{\thefigure}{S\arabic{figure}}
\setcounter{figure}{0}
\renewcommand{\thetable}{S\arabic{table}}
\setcounter{table}{0}

\section{Band touching points}

We give an exhaustive result for the properties of band touching points in this part. We illustrate how the exceptional points split and fuse into the Dirac points.

First we identify the nature of the band touching point. We focus on non-zero $G_\alpha$. The energy is given by $E(\mathbf q)=\pm\sqrt{A(\mathbf q)A(-\mathbf q)}$. In the Hermitian case, $A(-\mathbf q)=A^\ast(\mathbf q)$, the dispersion must be at least linear around the band touching point. In the non-Hermtian case, $A(-\mathbf q)\ne A^\ast(\mathbf q)$ in general. The dispersion is usually square-root dependent around the band-touching point. In fact, if $A(-\mathbf q)=A^\ast(\mathbf q)=0$, then we can find that $G_x\sin(\tilde q_1)=-G_y\sin(\tilde q_2)$. This requires either $\phi_x=\phi_y$ (mod $\pi$) or the band-touching point at the inversion-invariant point ($\tilde q_1,\tilde q_2=0,\pi$). In the first situation, for simplicity we can absorb the phases $\phi_x$ and $\phi_y$ into $\phi_z$. Using $\textrm{Im} A(\mathbf q)=\textrm{Im} A(-\mathbf q)=0$, we obtain $G_x\sin(\tilde q_1)+G_y\sin(\tilde q_2)+|G_z|\sin{\phi_z}=0$ and $-G_x\sin(\tilde q_1)-G_y\sin(\tilde q_2)+|G_z|\sin{\phi_z}=0$. So $\phi_z=0$ (mod $\pi$), all $G_\alpha$ have to share the same phase and the system is simply obtained by the Hermtian one times an overall phase. In the second situation where the band-touching points are at the inversion-invariant points, i.e. $\tilde q_1,\tilde q_2=0,\pi$, we as before absorb the phase of $G_z$ into the overall phase of the Hamiltonian. The band-touching condition is:
\begin{equation}
    |G_x|\cos(\tilde q_1+\bar\phi_x)+|G_y|\cos(\tilde q_2+\bar\phi_y)+G_z=0,\quad |G_x|\sin(\tilde q_1+\bar\phi_x)+|G_y|\sin(\tilde q_2+\bar\phi_y)=0 \label{eq_gpcri}.
\end{equation}
From these two equations, one finds that
\begin{equation}
    |G_x|=\frac{\sin(\bar\phi_y+\tilde q_2)}{\sin(\bar\phi_x+\tilde q_1-\bar\phi_y-\tilde q_2)} |G_z|,\quad |G_y|=\frac{\sin(\bar\phi_x+\tilde q_1)}{\sin(\bar\phi_y+\tilde q_2-\bar\phi_x-\tilde q_1)} |G_z|.
\end{equation}
This means we need a fine tuning to have the band-touching point located at the inversion-invariant point. Therefore, we conclude that the band-touching point is Dirac-like only for Hermitian Hamiltonians up to an overall complex phase or fine tuning where the band-touching point takes place at the inversion-invariant point in the Brillouin zone. Otherwise, the band-touching points possess a defect Hamiltonian and are exceptional points.

Then we consider the boundary between the gapless phase $B$ and the gapped phase $A$ when the Hamiltonian is non-Hermitian. We rewrite the band-touching condition as
\begin{align}
    &\cos(\tilde q_1+\bar\phi_x)=\frac{|G_y|^2-|G_z|^2-|G_x|^2}{2|G_x||G_z|},\quad
    \cos(\tilde q_2+\bar\phi_y)=\frac{|G_x|^2-|G_z|^2-|G_y|^2}{2|G_y||G_z|},\label{eq_spbtce}\\ &|G_x|\sin(\tilde q_1+\bar\phi_x)+|G_y|\sin(\tilde q_2+\bar\phi_y)=0.\label{eq_spbtc}
\end{align}
The boundary is reached when the solution to Eq.~\eqref{eq_spbtce} is at its extremes. That is, $|\cos(\tilde q_1+\bar\phi_x)|=1$ or $|\cos(\tilde q_2+\bar\phi_y)|=1$ in  Eq.~\eqref{eq_spbtce}.  According to Eq.~\eqref{eq_spbtc}, we actually have $|\cos(\tilde q_1+\bar\phi_x)|=|\cos(\tilde q_2+\bar\phi_y)|=1$. In this situation, there is only one solution to $A(\mathbf q)=0$. For the Hermitian situation, this means the two Dirac points appearing in pairs $\mathbf q_\ast, -\mathbf q_\ast$ are fusing into one. For general non-Hermitian case, according to the discussion in the above paragraph, the band-touching point $\mathbf q_e$ is usually not inversion invariant. So we obtain two exceptional points at the phase boundary $,\mathbf q_e,-\mathbf q_e$. We conclude that in general when the non-Hermitian system is evolving from the gappless to gapped phase, the four exceptional points fuse into two and then get gapped out.

We give the form of the Hamiltonian near the band-touching point. The Hamiltonian can be parameterized by the Pauli matrix as
\begin{equation}
    H=\frac{A(\mathbf q)+A(-\mathbf q)}{2}\sigma_x+\frac{A(\mathbf q)-A(-\mathbf q)}{2}i\sigma_y.
\end{equation}
For an Hermitian Hamiltonian, $A(-\mathbf q)=A^\ast(\mathbf q)$. At the Dirac point $\mathbf q_\ast$, we have $A(\mathbf q)\simeq g_\alpha \delta q_\alpha$, where $g_\alpha$ is a complex vector and $\delta \mathbf q=\mathbf q-\mathbf q_\ast$. So the Hamiltonian takes the form $H=\sigma_x \delta q_\alpha\textrm{Re } g_\alpha-\sigma_y\delta q_\alpha \textrm{Im } g_\alpha$. In general $\textrm{Re } g_\alpha$ and $\textrm{Im } g_\alpha$ are linearly independent. So the dispersion of the energy is linear $E\simeq \sqrt{g_\alpha g^\ast_\beta\delta q_\alpha\delta q_\beta}$. At the Hermitian phase boundary, using $\mathbf q_\ast=-\mathbf q_\ast$, one can deduce that $\textrm{Re } g_\alpha=0$ and $g_\alpha q_\alpha=0$. So the Hamiltonian becomes $H=ig_\alpha\delta_\alpha \sigma_y$ and $E\sim |\delta \mathbf q|^2$ for $\delta \mathbf q$ perpendicular to $\textrm{Im } g_\alpha$. In the non-Hermtian situation, at the band touching point only one of $A(\mathbf q),A(-\mathbf q)$ vanishes. Let's take $A(\mathbf q_e)=0$ and $A(\mathbf q)\simeq g_\alpha\delta q_\alpha$. Then the Hamiltonian and the energy are expressed as
\begin{equation}
    H\simeq \left(\begin{array}{ccc}
        0 & g_\alpha \delta q_\alpha\\
        A(-\mathbf q_e) & 0
    \end{array}\right), \quad E\simeq \sqrt{A(-\mathbf q_e)g_\alpha\delta q_\alpha}.
\end{equation}
As before, the real and the imaginary part of $g_\alpha$ are usually linearly independent. The energy is square-root dependent on $\delta q$. At the gap-gapless boundary, by examining the structure of Eq.~\eqref{eq_gpcri} as done in the last paragraph, the vector $g_\alpha$ is real up to a complex phase. In that case, along the orthogonal direction to $g_\alpha$, the dispersion $E(\mathbf q)$ is linear in $\delta q$.

Another interesting feature compared to the Hermitian situation is how the system can evolve from ferromagnetic couplings to anti-ferromagnetic couplings. We notice that the Majoranas are agnostic to whether the coupling is ferromagnetic or anti-ferromagnetic, and only depends on the modulus $|G_\alpha|$. However, in the Hermitian case, we have to pass through the $G_\alpha=0$ point, which is equivalent to an array of $1$D gapless chains. Below we list how the exceptional points and the Fermi arcs evolves when $\phi_z$ going from $0$ to $\pi$ at fixed $G_x=2,G_y=1$ in Fig.~\ref{fig_FAEvo}. From ferromagnetic $G_z$ coupling to anti-ferromagnetic $G_z$ coupling, one can observe that there are two recombination processes of the Fermi arc. Two Fermi arcs appear in splitting each Dirac point to two exceptional points. Between $\phi_z=\pi/4$ and $\phi_z=5\pi/12$, the two Fermi arcs collide for the first time. How they connect the four exceptional points into two pairs changes. Each Fermi arc together with their imaginary counterpart winds over the torus hole, forming a closed path that can not shrink into a point. Between $\phi_z=7\pi/12$ to $\phi_z=3\pi/4$, the Fermi arcs collide again. This time the Fermi arcs again connect the exceptional points in the same way as for small $\phi_z$. And they eventually diminish when $\phi_z$ is approaching $\pi$. In summary, in the non-Hermitian situation, we can go from a anti-ferromagnetic Kitaev spin liquid $J_\alpha<0$ to an ferromagnetic Kitaev spin liquid $J_\alpha>0$ by splitting and reconnecting exceptional points. While in the Hermitian case, one has to go through the critical regime $J_\alpha=0$ exhibiting nodal lines. The non-Hermitian coupling thus provides paths circumventing this critical point.

\begin{figure}
\centering
     \subfloat[]{
    \includegraphics[width=0.32\linewidth]{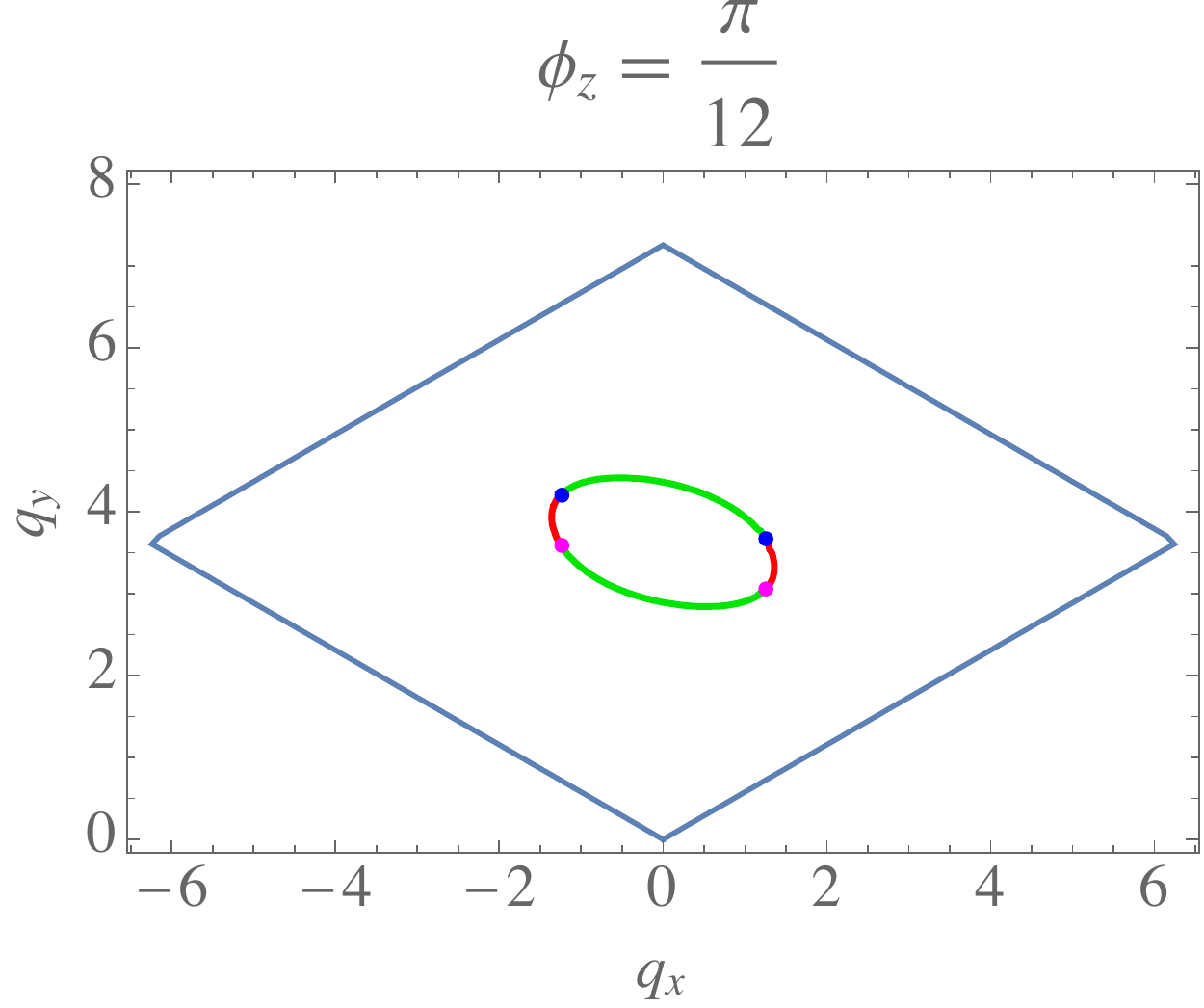}
    }
     \subfloat[]{
    \includegraphics[width=0.32\linewidth]{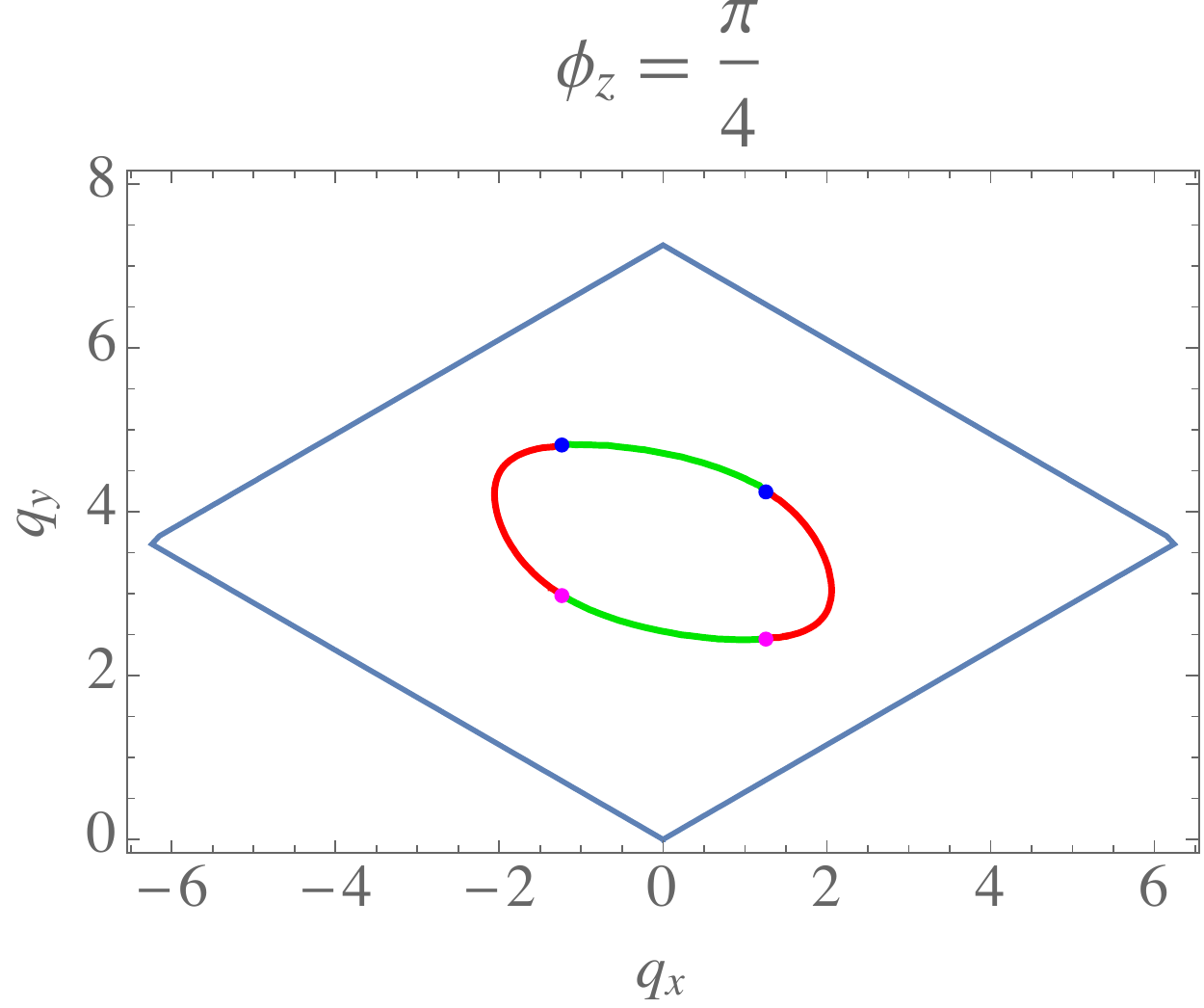}
    }
      \subfloat[]{
    \includegraphics[width=0.32\linewidth]{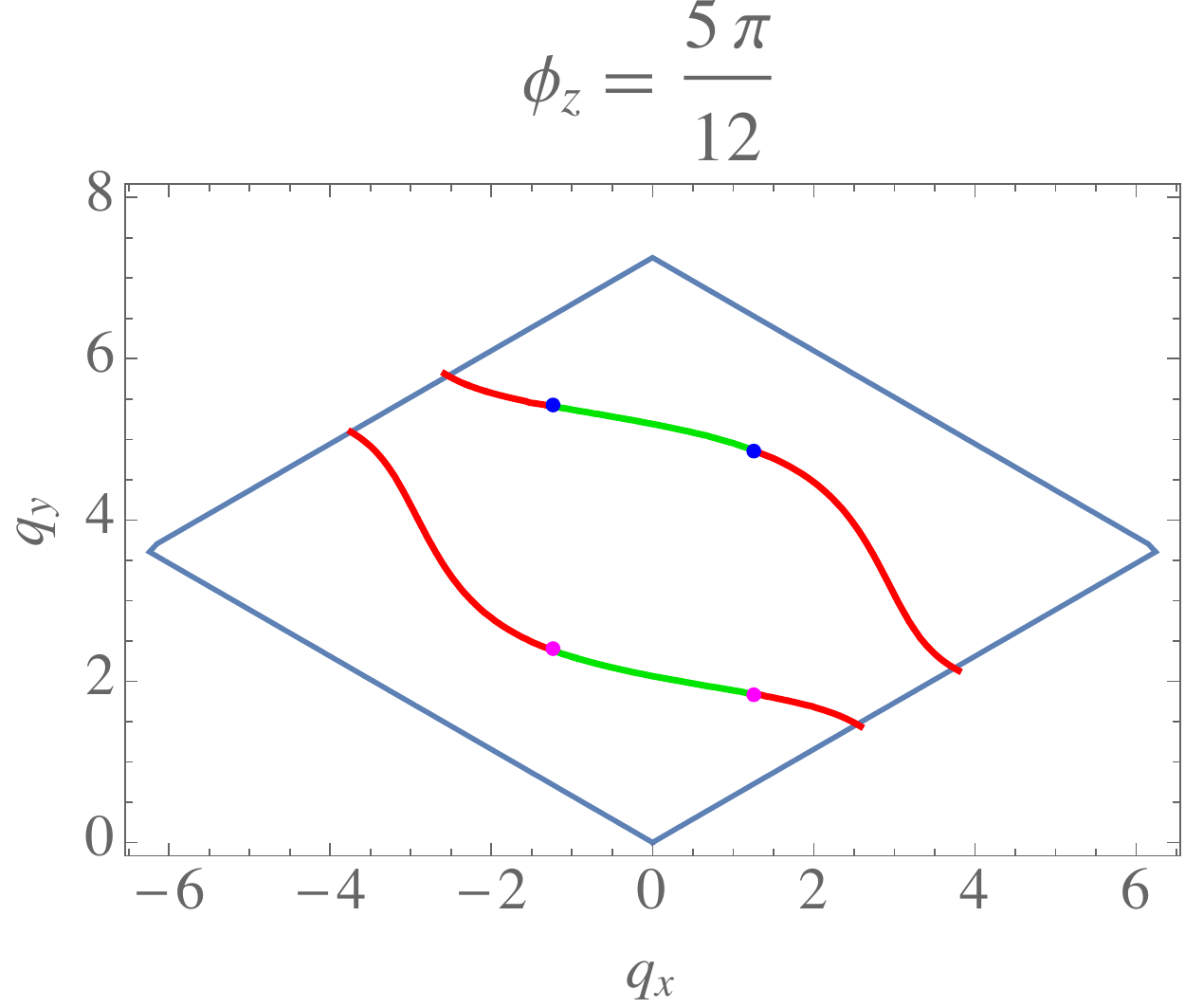}
    }
    
     \subfloat[]{
    \includegraphics[width=0.32\linewidth]{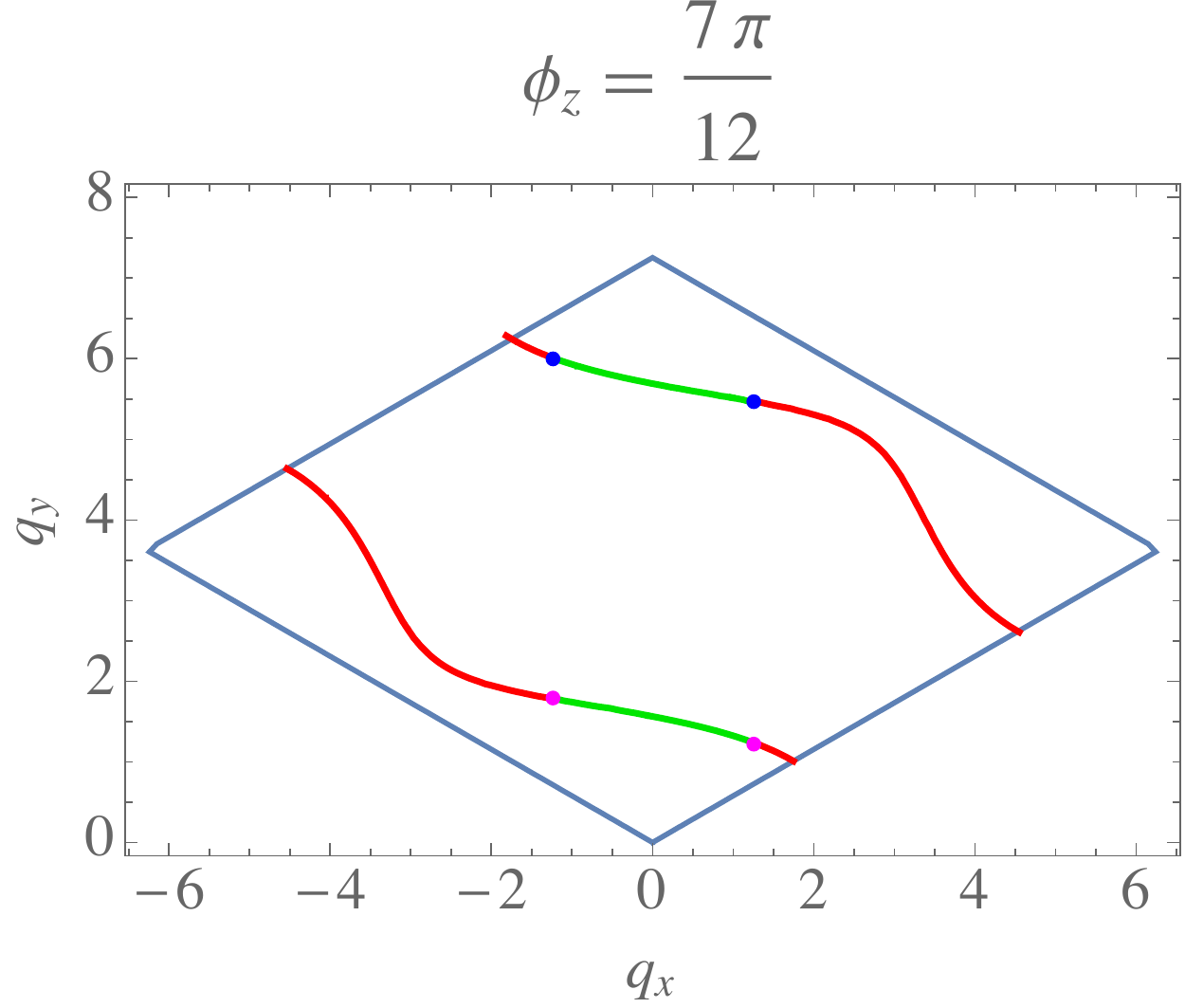}
   }
     \subfloat[]{
    \includegraphics[width=0.32\linewidth]{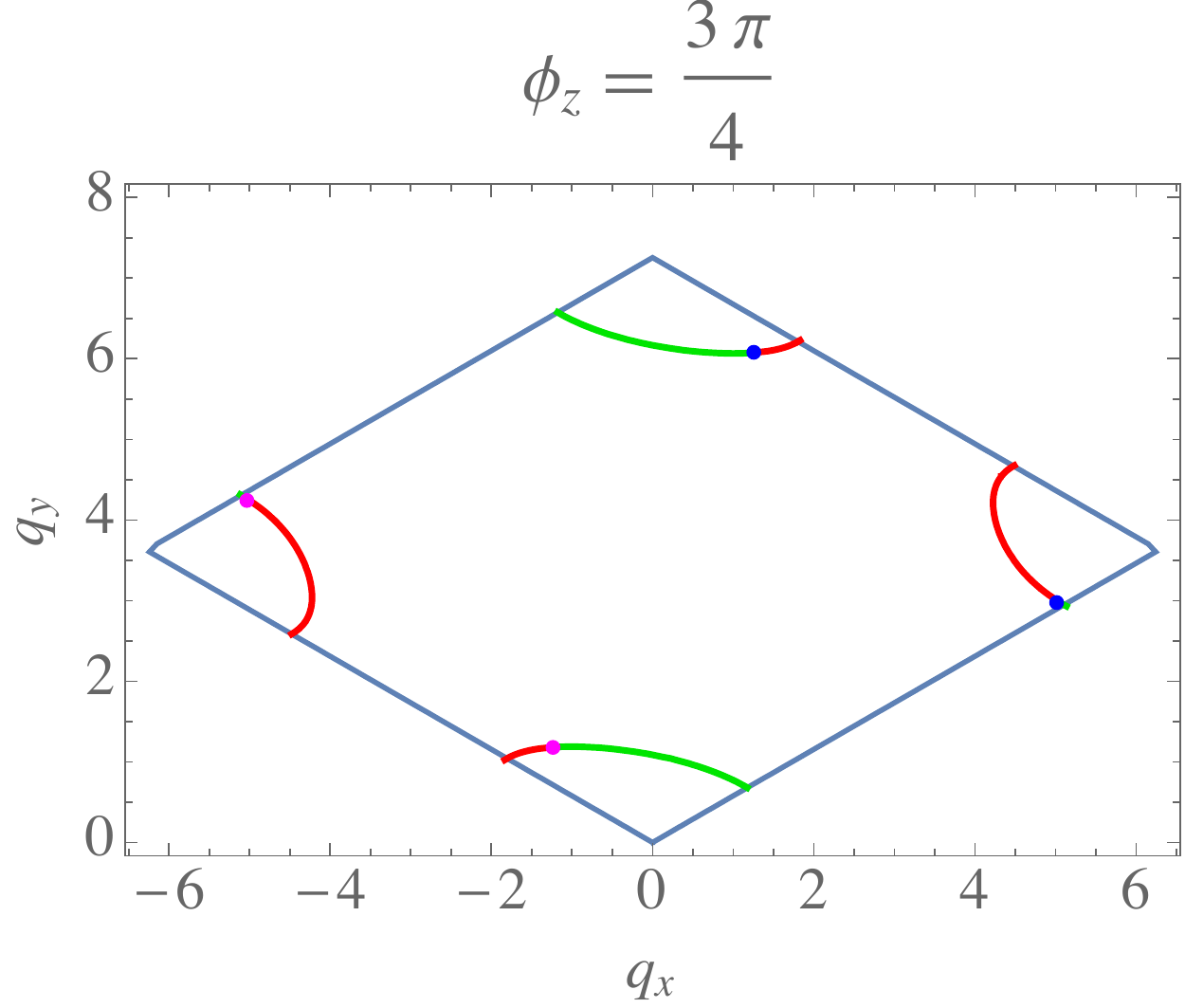}
    }
      \subfloat[]{
    \includegraphics[width=0.32\linewidth]{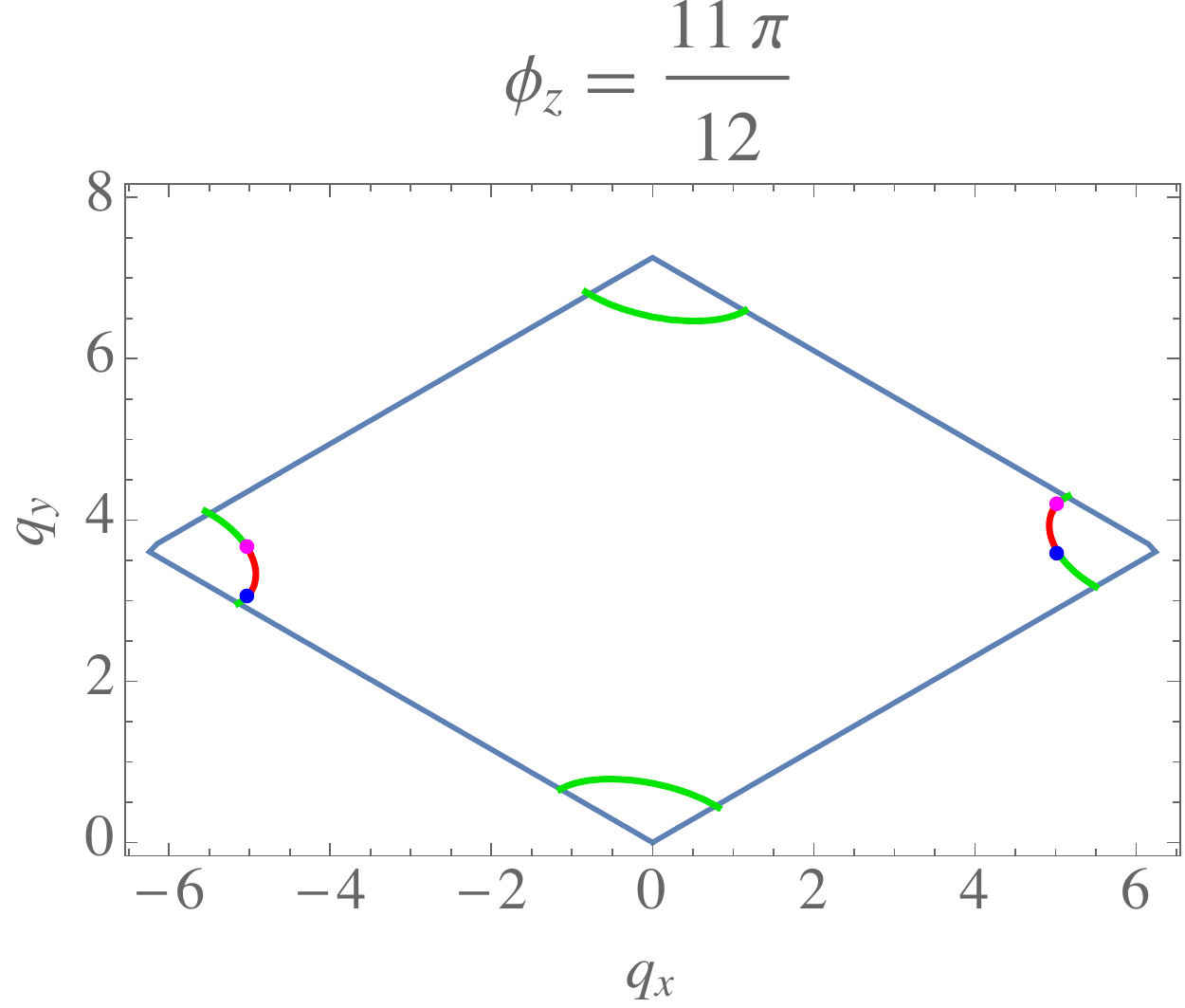}
    }
    \caption{(a)-(f) The evolution of the band-touching points, Fermi arcs (red) and the $\textrm{Im} E=0$ curve (green) for $G_x=2, G_y=1$ and $G_z=2.5\exp(i\phi_z)$ from $\phi_z=\pi/12$ to $\phi_z=11\pi/12$.}
    \label{fig_FAEvo}
\end{figure}

\section{The transfer matrix solution to open boundary condition}

Now we solve the OBC problem. The Hamiltonian is diagonal in momentum $q_x$. We do the Fourier transformation of the Majorana operators as $c_{q_x,m}=\sum_n (1/\sqrt{2N_x})c_{n,m}e^{-iq_x n}$. The Hamiltonian then takes the form $\widetilde H=\sum_{m,m',q_x}A^O_{m,n}(q_x)c_{-q_x,m}c_{q_x,m'}$, where the matrix $A^O$ is given by
\begin{equation}
    A^O=\begin{pmatrix}
         0 & ir(q_x) &\ &\ &\ &\ &\ \\
         -ir'(q_x) & 0 &it &\ &\ &\ &\ \\
         \  &-it' & 0 & ir(q_x) &\  &\ &\ \\
         \ &\  &-ir'(q_x) &0&it &\ &\ \\
         \ &\ &\ &\ddots &\ddots &\ddots &\ 
    \end{pmatrix},
\end{equation}
where $r(q_x)=r'(-q_x)=2\left[G_xe^{-iq_xa/2}+G_ye^{iq_xa/2}\right]$ and $t=t'=-2G_z$.

\begin{figure}
     \centering
     \subfloat[]{
     \includegraphics[width=0.32\linewidth]{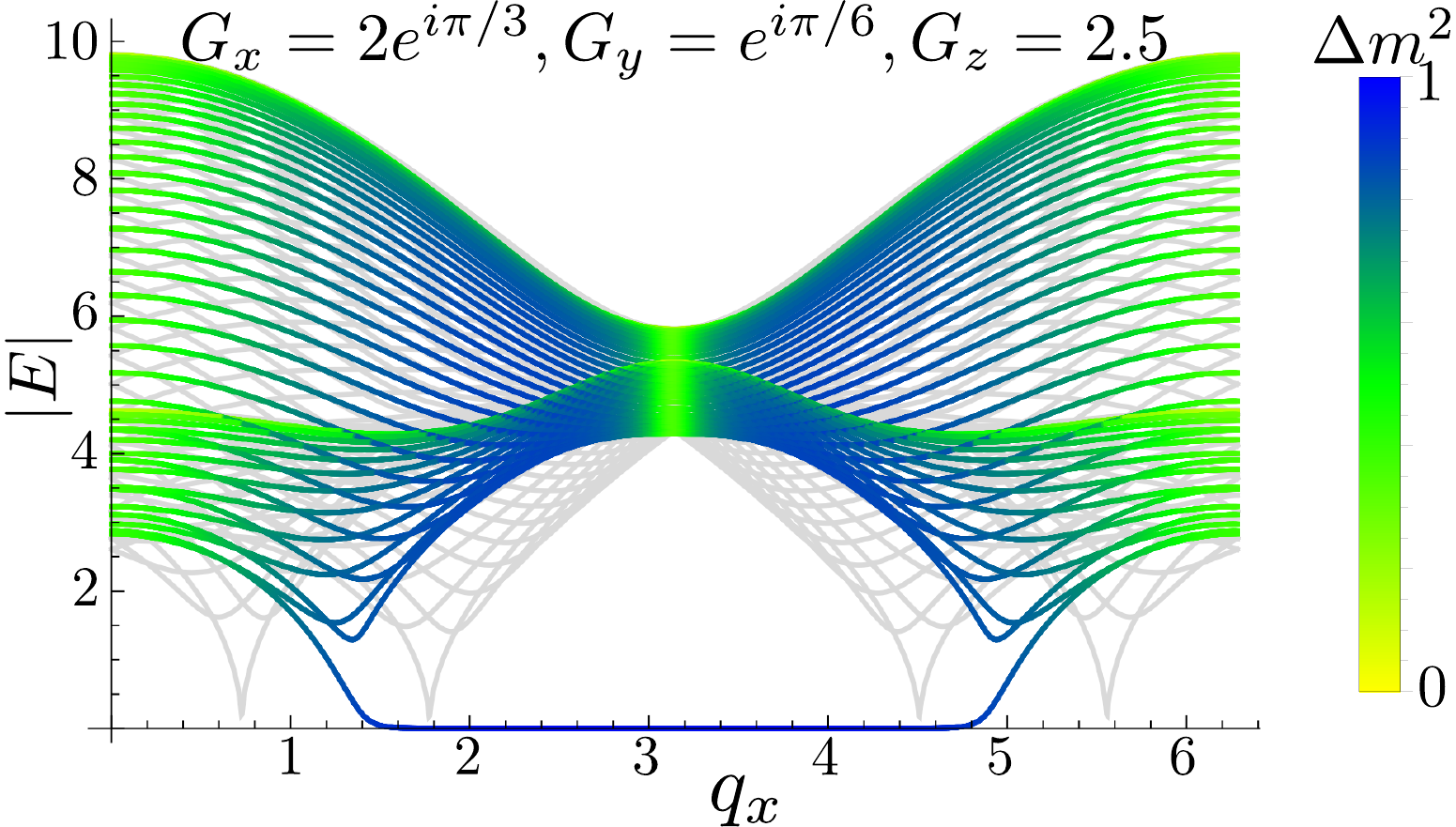}
    \label{fig_opbspecv}
    }
    \subfloat[]{
    \includegraphics[width=0.32\linewidth]{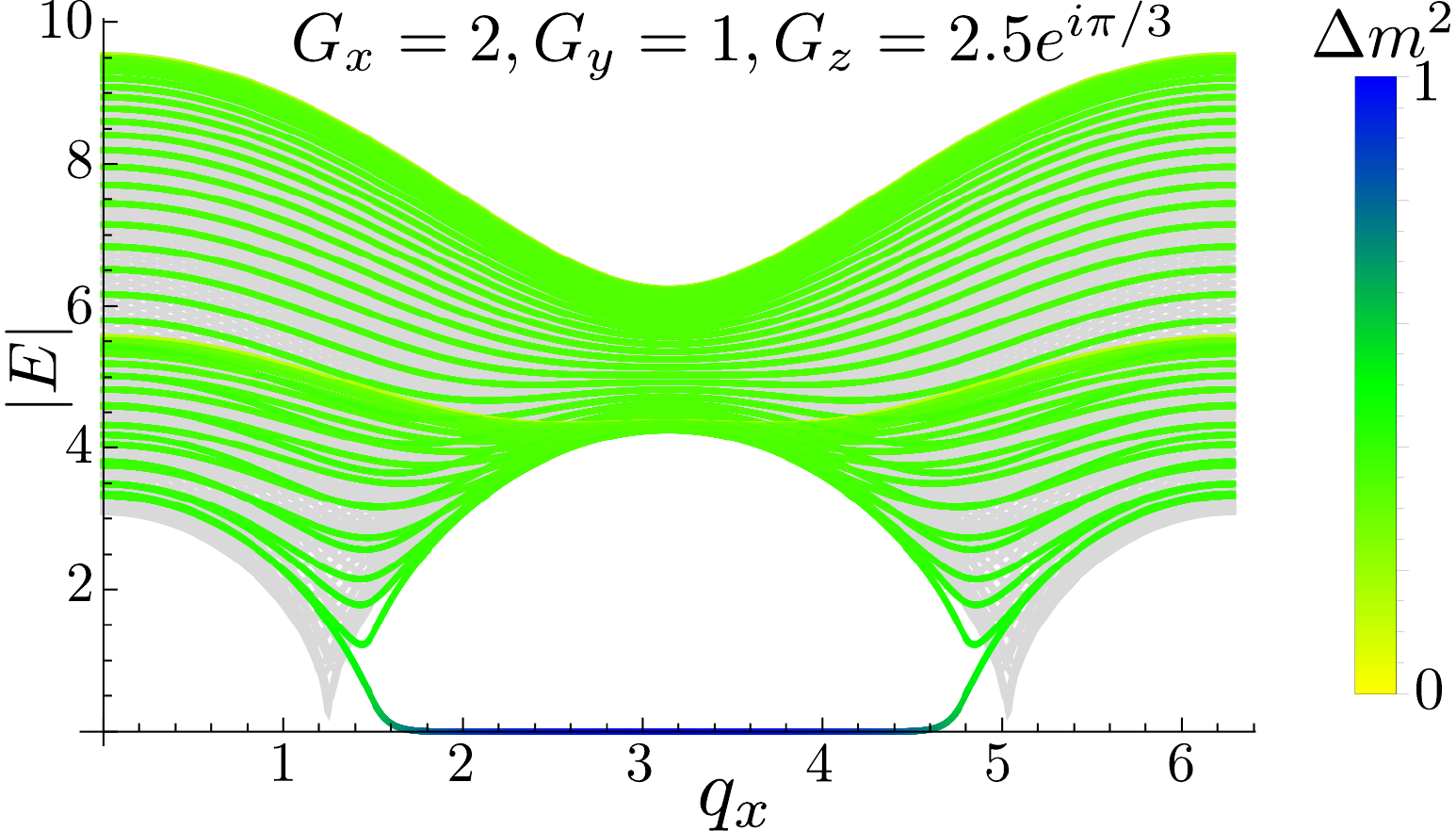}
    \label{fig_skineffNv}
    }
     \subfloat[]{
    \includegraphics[width=0.32\linewidth]{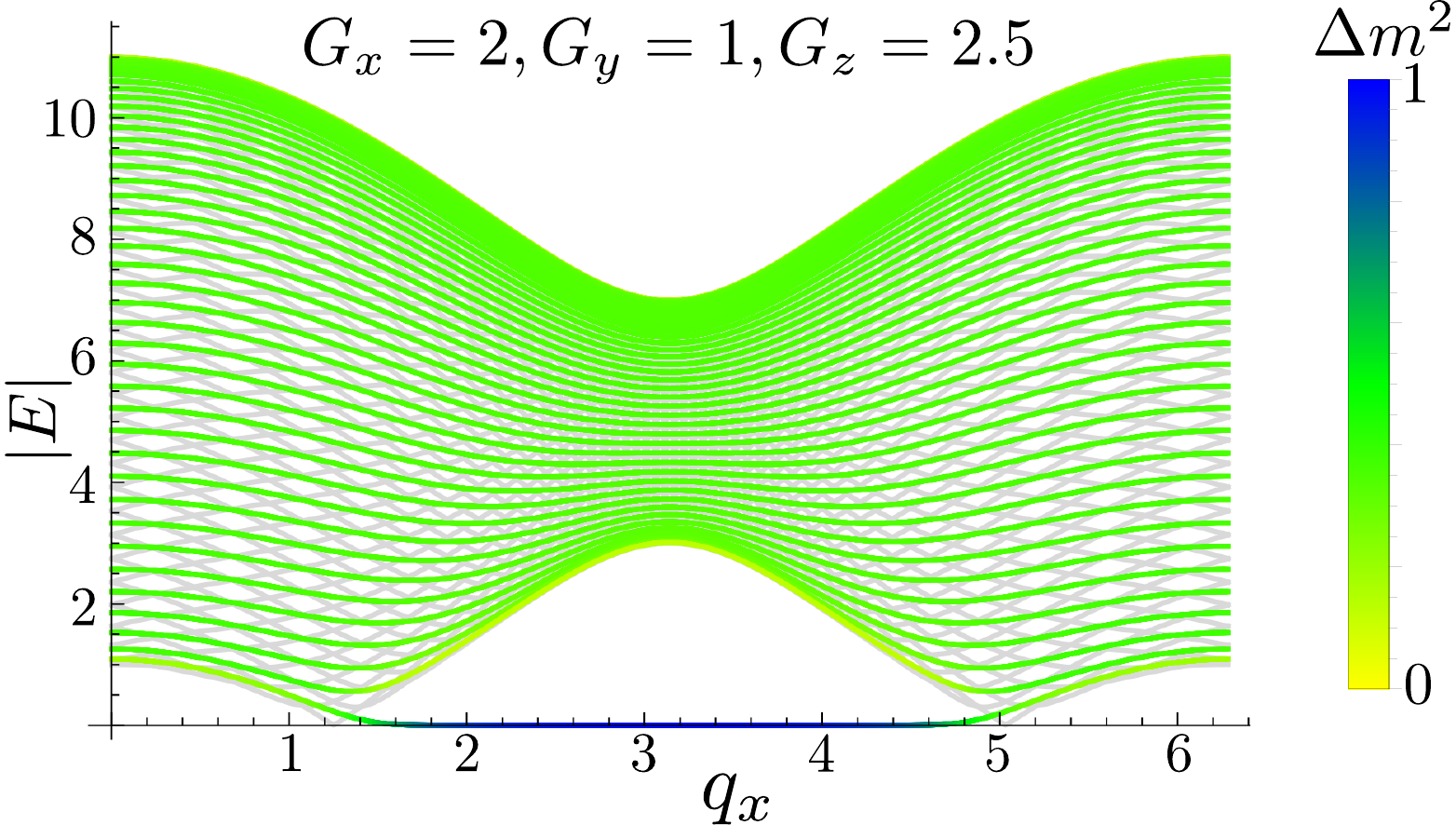}
    \label{fig_skineffv}
    }
    
     \hspace{0.027\linewidth}
      \subfloat[]{
     \includegraphics[width=0.26\linewidth]{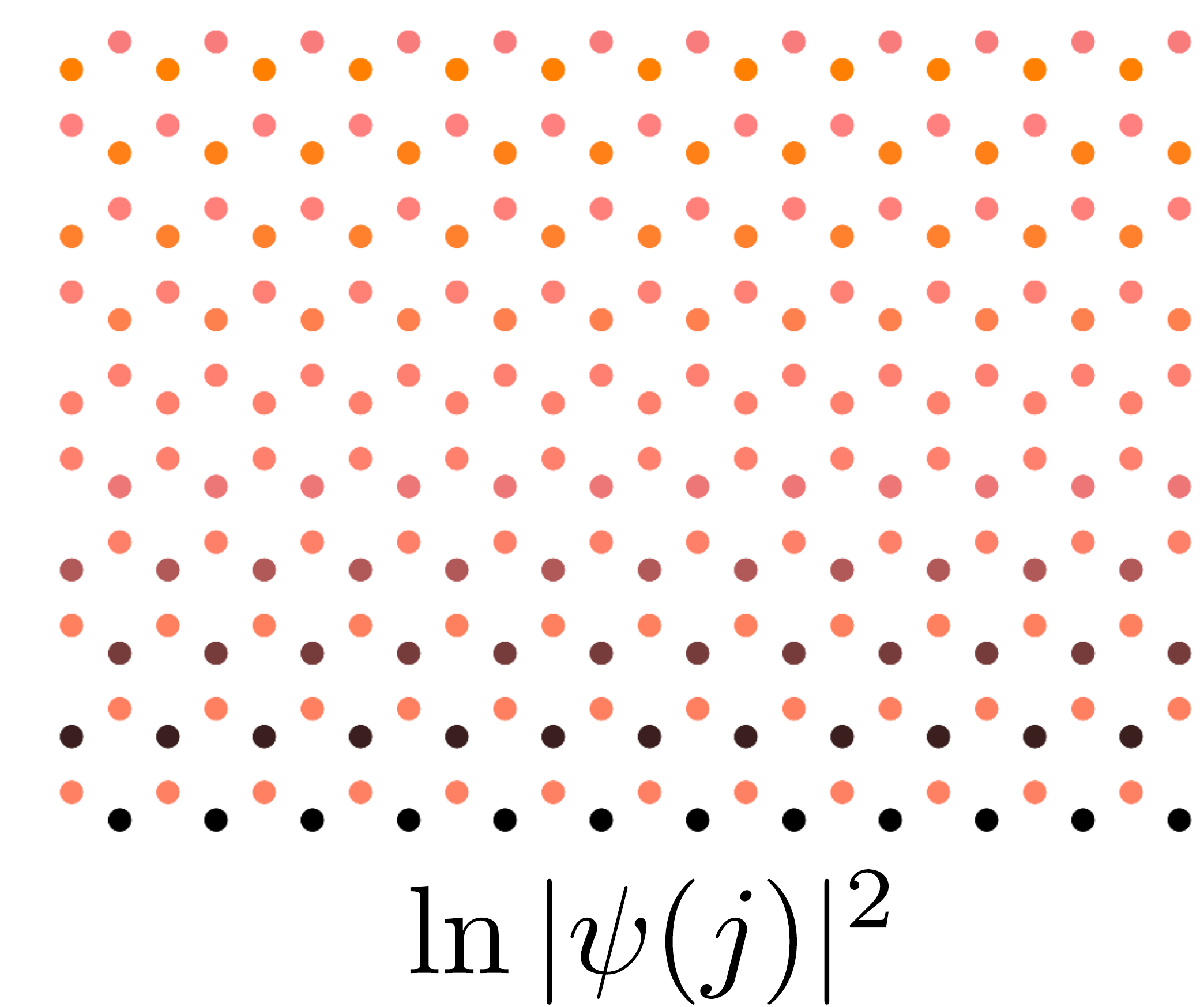}
    \label{fig_skltc}
    }\hspace{0.053\linewidth}
    \subfloat[]{
     \includegraphics[width=0.26\linewidth]{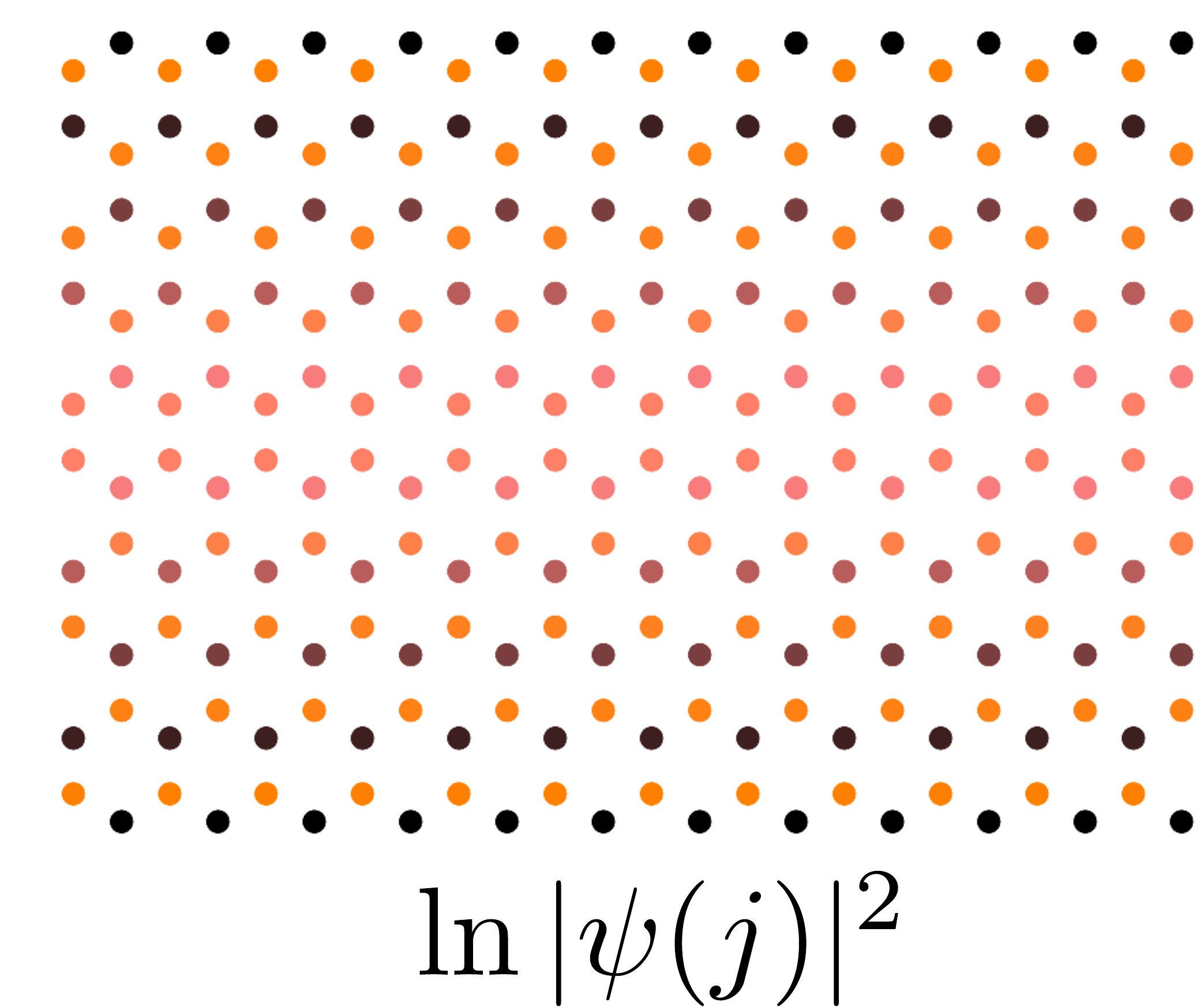}
    \label{fig_nskltc}
   }\hspace{0.027\linewidth}
     \subfloat[]{
    \centering
     \includegraphics[width=0.32\linewidth]{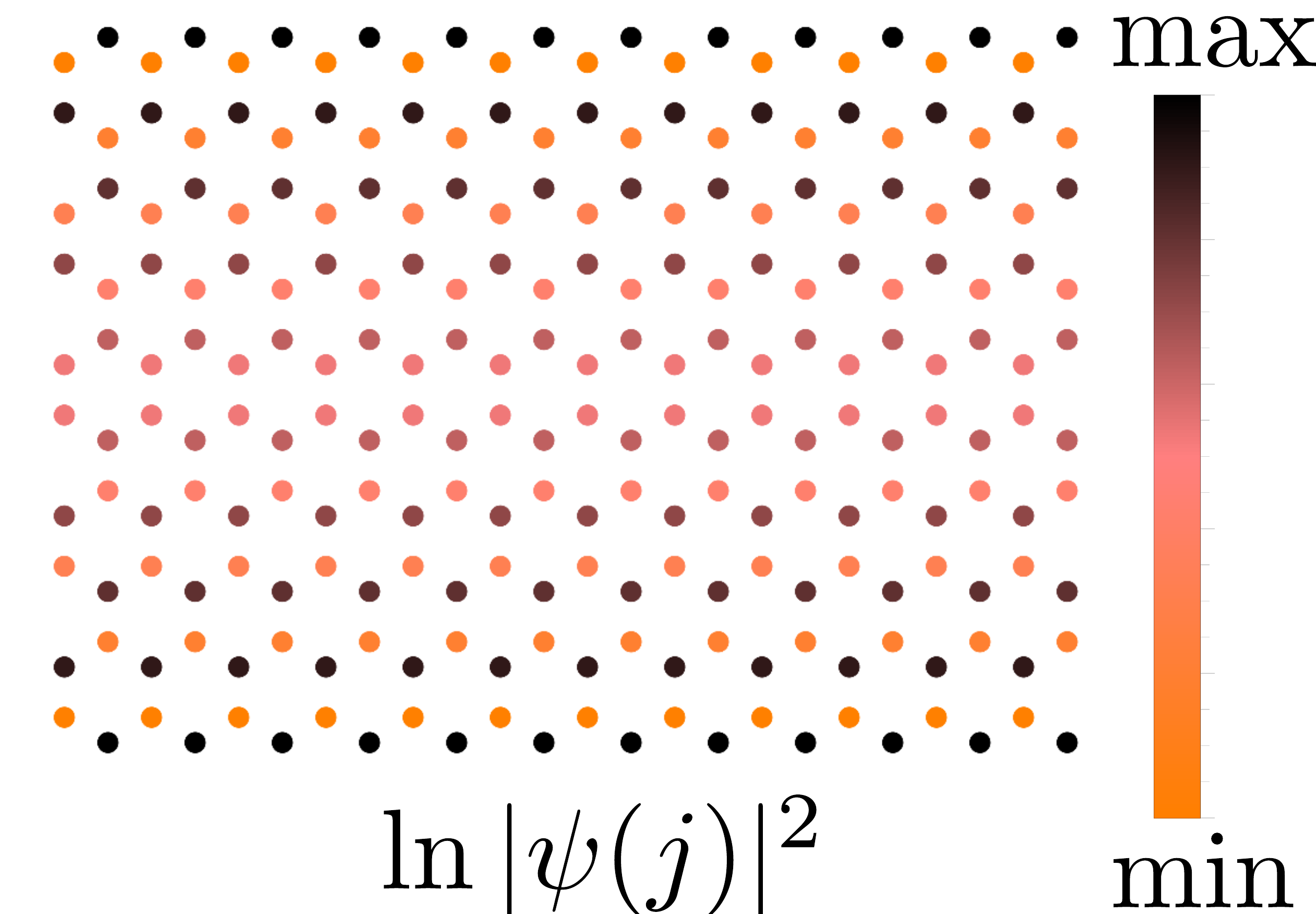}
    \label{fig_hltc}
   }
    \caption{The spectra for OBC (colored) and  PBC (light gray) and the corresponding localisation variance of the state. The calculation is performed on a $M=80$-row lattice (a) The skin effect situation. (b) No skin effect exhibits when only $G_z$ is tuned complex. The zero-mode edge state is performed on a slightly smaller OBC lattice. (c) The Hermitian result, also with the zero-mode edge computed on a slightly smaller OPC lattice. (d)-(f) The corresponding logarithmic distribution of the wave functions $\ln |\psi(j)\psi(j)|$ on the lattice at $q_x=2\pi/3$.}\label{spple_plot}
\end{figure}

We use the transfer matrix to analyze the eigenstates $\psi(m)$ of the above matrix. Like in the main text, we rewrite them as a doublet $\Psi(m)=(\psi(2m),\psi(2m+1))$. The eigenstate equation becomes 
\begin{equation}
    T_+\Psi(m)+T_-\Psi(m-1)=0,\quad \Rightarrow \Psi(m)=T\Psi(m-1),
\end{equation}
where those transfer matrices are given by
\begin{equation}
    T_+=\left(\begin{array}{ccc}
        ir & 0\\
        -E & it
    \end{array}\right),\quad 
    T_-=\left(\begin{array}{ccc}
        -it' & -E\\
        0 & -ir'
    \end{array}\right),\quad T=-T_+^{-1}T_-=\frac{1}{tr}\left(\begin{array}{ccc}
        tt' & -itE\\
        -it'E & rr'-E^2
    \end{array}\right).
\end{equation}
In order to find the solution for OPB, we need to compute $T^{M/2}$, which is convenient by transforming $T$ into a diagonal matrix through a similarity transformation:
\begin{equation}
    T=P\left(\begin{array}{ccc}
        s_1 & 0 \\
         0 & s_2
    \end{array}\right)P^{-1},\quad 
    s_1,s_2=\frac{1}{2tr}\left[rr'+tt'-E^2\mp \sqrt{(rr'-tt'-E^2)^2-4tt'E^2}\right].
\end{equation}
where $P$ is formed by the columns of the right eigenvectors of $T$ and $P^{-1}$ is formed by the rows of the left eigenvectors of $T$. The boundary condition now can be written as:
\begin{equation}
    P\left(\begin{array}{ccc}
        s_1^{M/2} & 0 \\
         0 & s_2^{M/2}
    \end{array}\right)P^{-1}\left(\begin{array}{ccc}
         0 \\
         \psi(1) 
    \end{array}\right)=\left(\begin{array}{ccc}
         \psi(M) \\
         0 
    \end{array}\right)\Rightarrow  \left(\begin{array}{ccc}
        A\left(s_1^{M/2}-s_1^{M/2}\right) \\
        Bs_1^{M/2}+(1-B)s_2^{M/2}
    \end{array}\right)\psi(1)=\left(\begin{array}{ccc}
         \psi(M) \\
         0 
    \end{array}\right),
\end{equation}
where the values $A,B$ are given by
\begin{equation}
    A=\frac{it}{\sqrt{(rr'-tt'-E^2)^2-4tt'E^2}},\quad B=\frac{1}{2}-\frac{rr'-tt'-E^2}{2\sqrt{(rr'-tt'-E^2)^2-4tt'E^2}}.
\end{equation}
The energy is solved by the implicit equation $ Bs_1^{M/2}+(1-B)s_2^{M/2}=0$. In the large-$M$ limit, $|s_1/s_2|=|(1-B)/B|^{2/M}$. This value goes to $1$ if neither $B$ nor $1-B$ vanishes, which requires $E\ne 0$. Therefore for a state with finite $|E|$, we can deduce $|s_1|\simeq |s_2|$. The distribution of the bulk state is thus derived as in the main text as $|s_1|\simeq |s_2|\simeq \sqrt{|\det T|}$. In Ref. \cite{PhysRevB.99.245116}, the authors show that in order to obtain a dense spectrum in the $M\to\infty$ limit, one should have $|s_1|=|s_2|$. States with $|s_1|\ne|s_2|$ are exceptional and treated as boundary states.

In figure \ref{spple_plot}, we plot the average square position $\Delta m^2=\sum [m-(M+1)/2]^2|\psi(m)|^2$ and the distribution of the wave function on the real-space lattice.

\section{Two-vortex sector}

\begin{figure}
    \centering
    \includegraphics[width=0.9\linewidth]{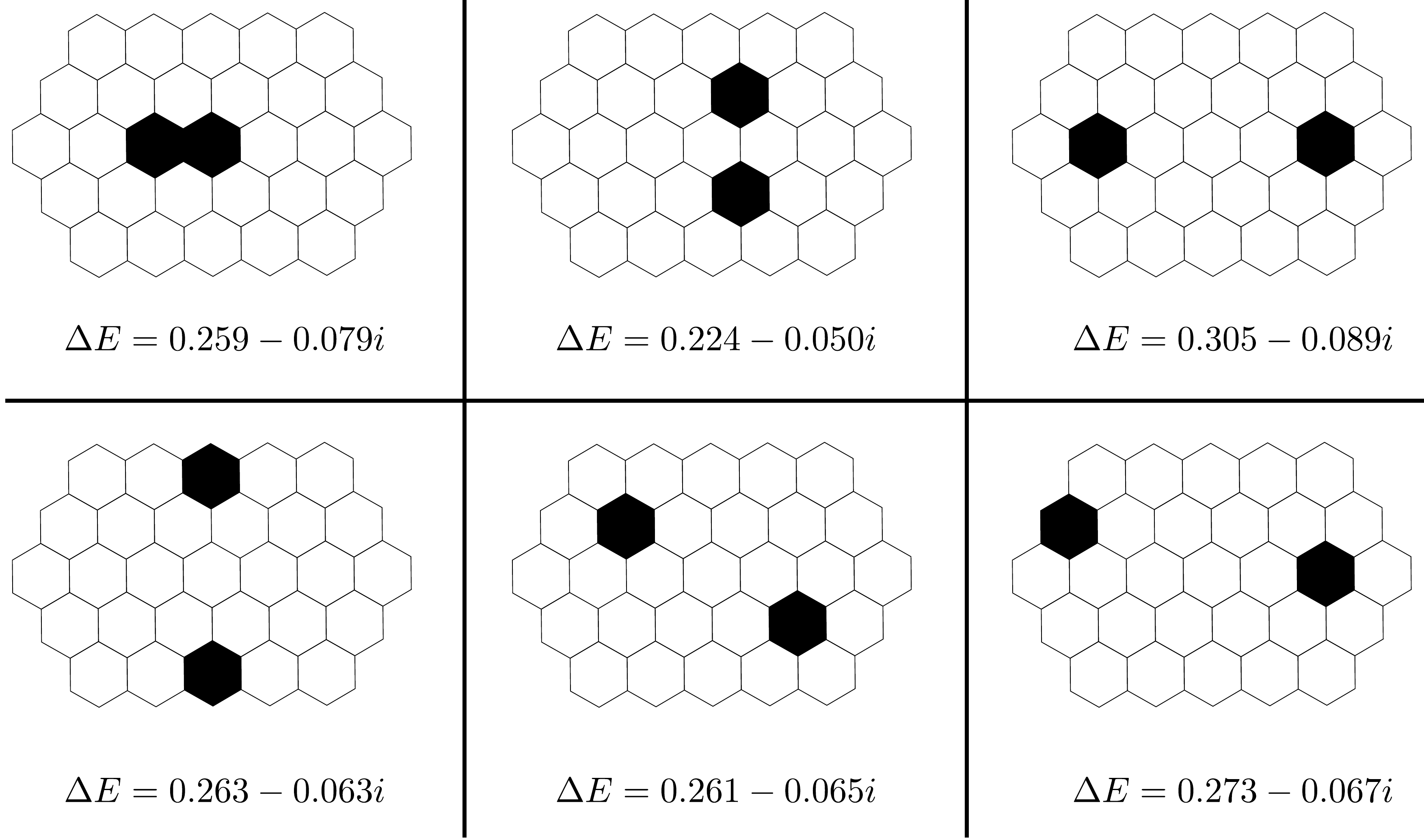}
    \caption{The energy of a series of two-vortex configurations at $J_x=J_y=\exp(-0.2i), J_z=\exp(-0.3i)$. }
    \label{fig_e2v}
\end{figure}

We numerically compute the energy of the two-vortex sector to demonstrate that it is possible to have the zero-flux sector equipped with a longer lifetime.

In order to realize this, we first put an overall phase to the Hamiltonian so that the zero-flux sector has the longest lifetime. Then we add a non-trivial non-Hermitian perturbation. The calculation is performed on a torus of $50\times 50$ unit cells with periodic boundary condition.

Inside each vortex sector, there are a macroscopic number of states. When the Hamiltonian is Hermitian, the most relevant state in each sector is the one with the lowest energy. For a non-Hermitian Hamiltonian, usually we have a state with the lowest energy and a different state with the longest lifetime. If we consider a non-Hermitian perturbation and a short-time evolution compared to the lifetime, the most relevant procedures come from the excitation to higer-energy states. In this case, it is reasonable to consider the lowest-energy state inside each vortex sector. The energy of the two-vortex configuration is then given by summing all the Majorana states with negative $\textrm{Re} E$. This gives the usual definition of the ground state, the state with lowest $\textrm{Re} E$. One difference here is that finite-size effects can be very large. This is because it possible to have Majorana states with small $\textrm{Re} E$ but big $\textrm{Im}E$. They lead to a large oscillation in the imaginary part of the two-vortex energy. In order to justify a finite-size calculation, we have to limit ourselves to very small non-Hermitian couplings. In our tentative numerical calculation, we find that only at very extremely small non-Hermitian perturbations, $\phi_\alpha<0.05$, can the finite-size effect be completely removed. This leads to very small separation between EPs. Nevertheless, for a slightly larger $\phi_\alpha$, we still have a zero-flux sector with a longer lifetime despite finite-size effects.

We compute the all two-vortex configurations on a $25\times 25$ quarter of the torus at $J_x=J_y=\exp(-0.2i), J_z=\exp(-0.3i)$. We verified that all these configurations have a shorter lifetime and higher energy than the vortex-free configuration. Some of the configuration are presented in Fig.~\ref{fig_e2v}. The energy difference between the two-vortex configuration and the zero-vortex sector has a positive $\textrm{Re} \Delta E$ so that the zero-vortex sector is the ground state. The imaginary part $\textrm{Im}\Delta E$ is negative. This guarantees that $\exp(-i\Delta Et)$ is a decaying factor and all two-vortex configurations decays faster than the zero-vortex sector. In Fig.~\ref{fig_evl}, we list the energy of the two-vortex configuration with respect to the vortex distance. One see for $J_x=J_y=\exp(-0.2i), J_z=\exp(-0.3i)$, there is still a jump in $\textrm{Im }E$. Such an irregularity disappears for $J_x=J_y=\exp(-0.05i), J_z=\exp(-0.1i)$

\begin{figure}
     \centering
     \subfloat[]{
     \includegraphics[width=0.47\linewidth]{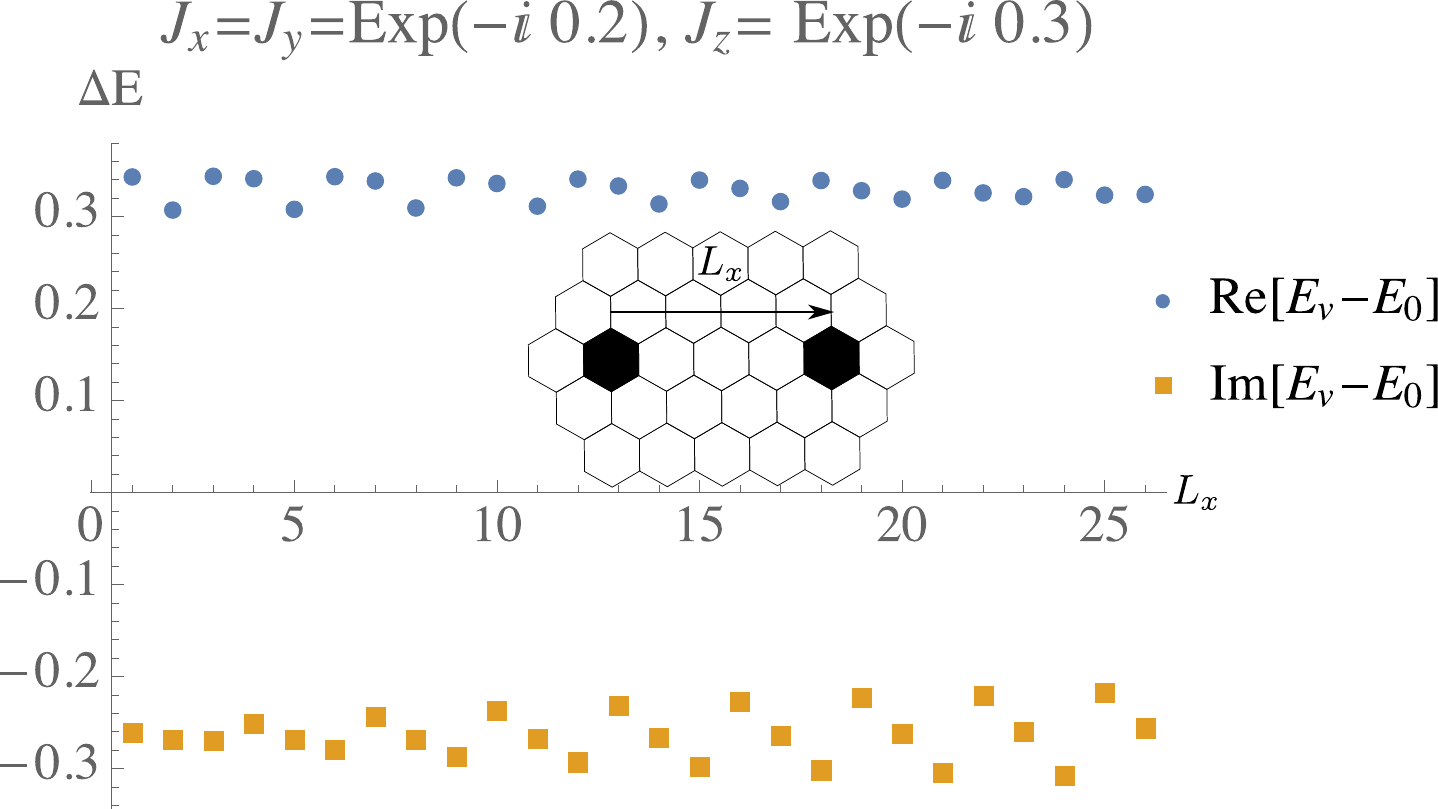}
    \label{fig_evx01}
    }
    \subfloat[]{
    \includegraphics[width=0.47\linewidth]{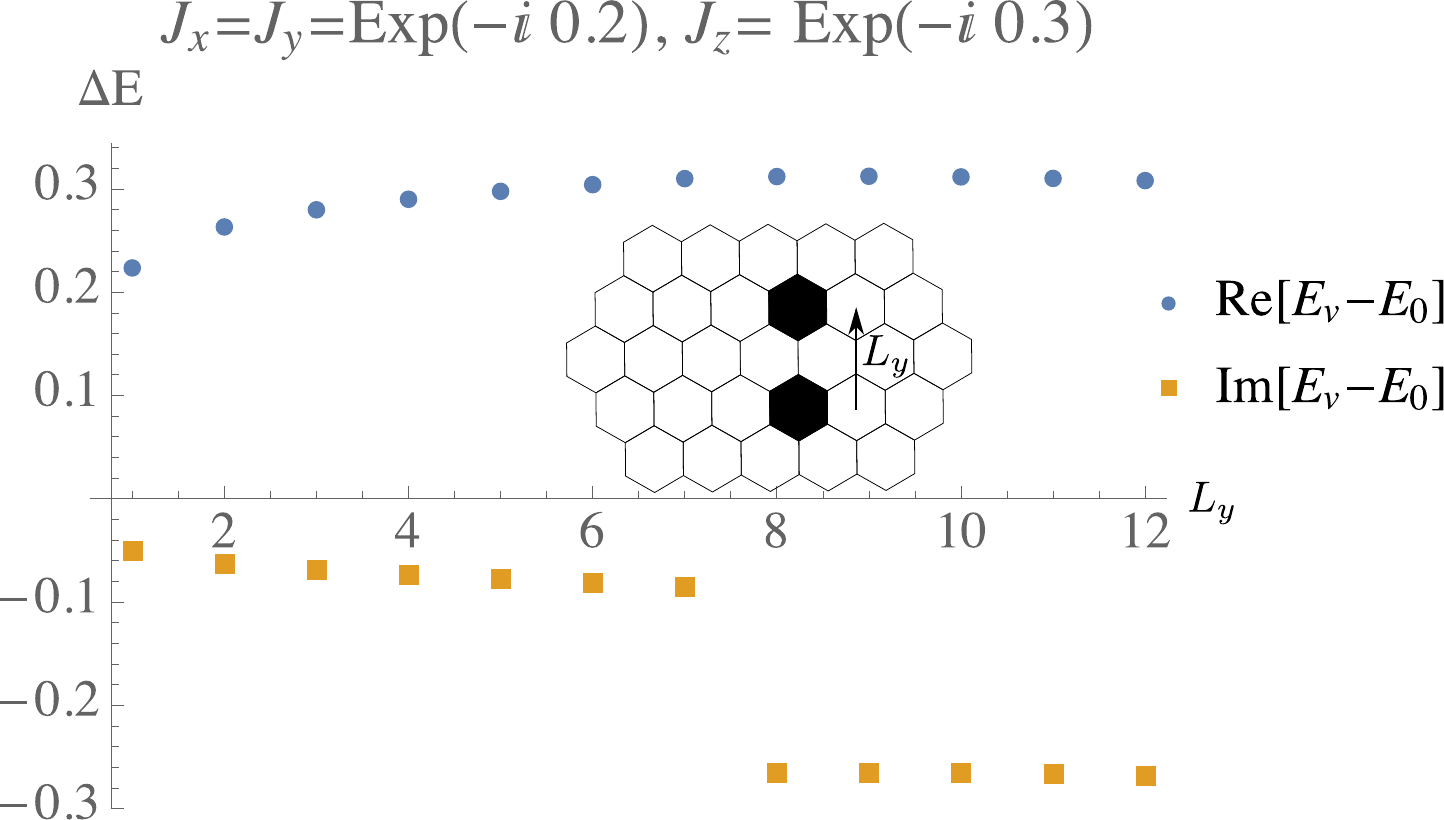}
    \label{fig_evy01}
    }
    
    \subfloat[]{
     \includegraphics[width=0.47\linewidth]{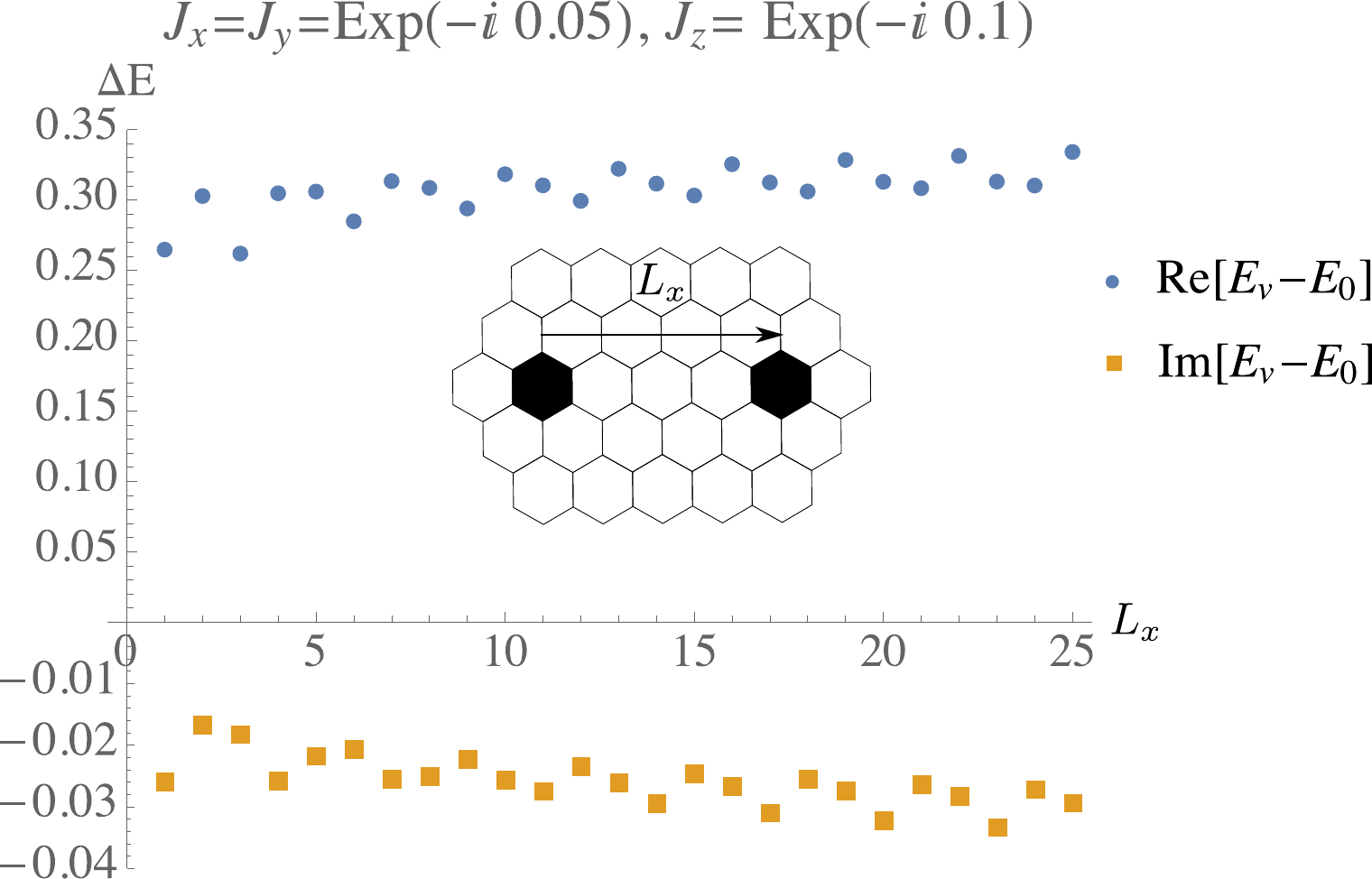}
    \label{fig_evx001}
    }
    \subfloat[]{
    \includegraphics[width=0.47\linewidth]{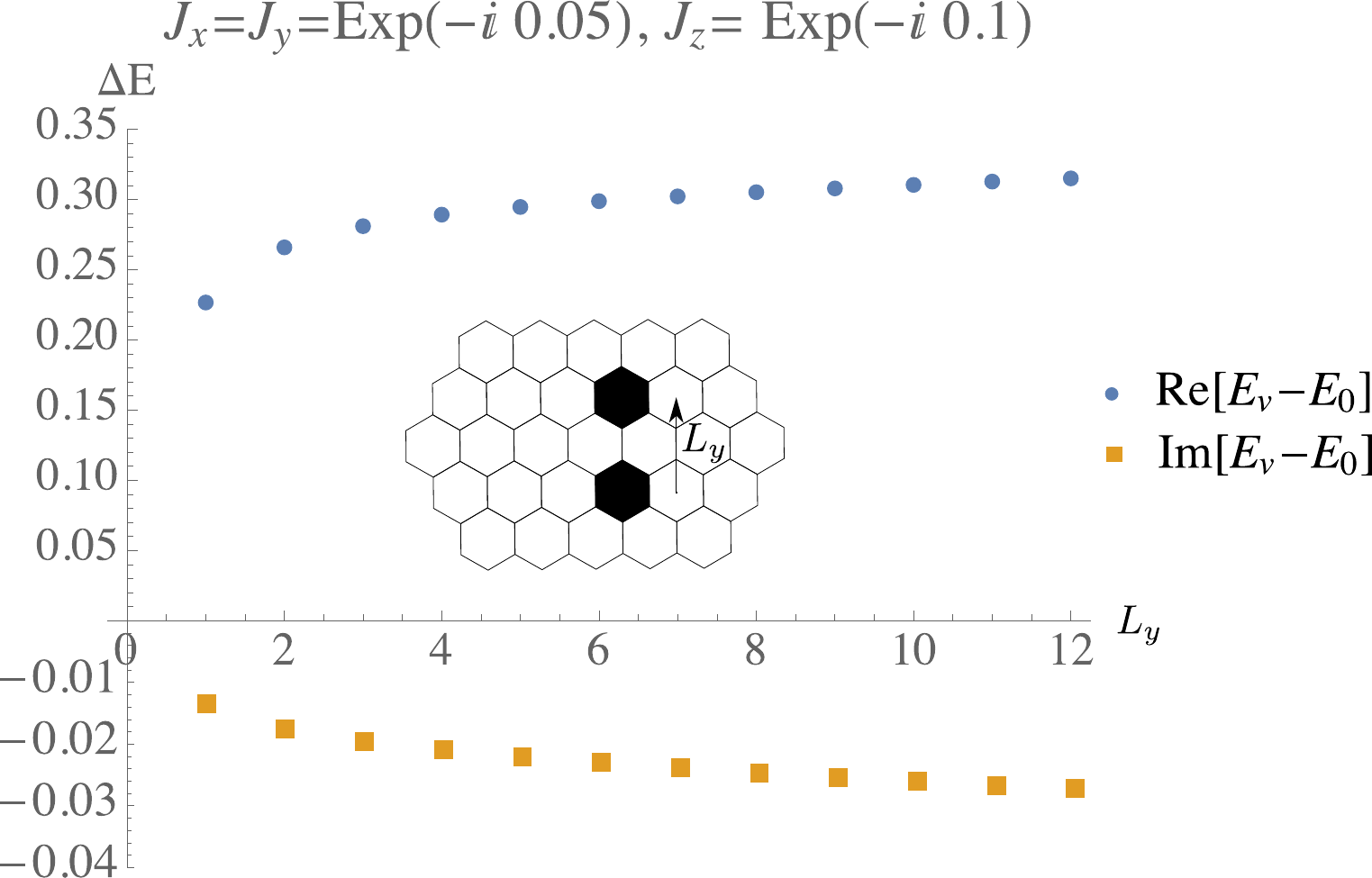}
    \label{fig_evy001}
    }
    \caption{The energy of the two-vortex configuration as a function of the vortex distance $L_x,L_y$.}\label{fig_evl}
\end{figure}

\end{widetext}

\end{document}